\documentclass[aps,pra,twocolumn,showpacs,amsmath,amssymb,superscriptaddress]{revtex4-1}
\usepackage[utf8]{inputenc}

\usepackage{tikz}
\usepackage{graphicx}
\usepackage{amsfonts}
\usepackage{amsmath}
\usepackage{amssymb,bbold}
\usepackage{color}
\usepackage{bm}
\usepackage{bbm}
\usepackage{dsfont}
\usepackage{enumerate}
\usepackage{amsfonts, amsmath, amsthm, amssymb} 
\usepackage{mathtools}
\usepackage{array}
\usepackage{mathdots}
\usepackage[colorlinks,bookmarks=false,citecolor=blue,linkcolor=red,urlcolor=blue]{hyperref}

\def\bra#1{\mathinner{\langle{#1}|}}
\def\ket#1{\mathinner{|{#1}\rangle}}
\newcommand{\gm}[1]{\textcolor{blue}{#1}}

\begin{document}

\title{Rise and fall of entanglement between two qubits in a non-Markovian bath}
\author{Sayan Roy}
\affiliation{Theoretische  Physik,  Universit\"at  des  Saarlandes,  D-66123  Saarbr\"ucken,  Germany}
\author{Christian Otto}
\affiliation{Theoretische  Physik,  Universit\"at  des  Saarlandes,  D-66123  Saarbr\"ucken,  Germany}
\author{Rapha\"el Menu}
\affiliation{Theoretische  Physik,  Universit\"at  des  Saarlandes,  D-66123  Saarbr\"ucken,  Germany}
\author{Giovanna Morigi}
\affiliation{Theoretische  Physik,  Universit\"at  des  Saarlandes,  D-66123  Saarbr\"ucken,  Germany}
\date{\today}


\begin{abstract}
We analyse the dynamics of quantum correlations between two qubits coupled to a linear chain of oscillators. The {chain mediates interactions between the qubits and acts as a} non-Markovian reservoir. The the model is amenable to an analytical solution when {the initial state of the chain is Gaussian}. We study the dynamics of the qubits concurrence starting from a separable state and assuming that the chain spectrum is gapped {and the chain is initially in a thermal state}. We identify three relevant regimes that depend on the strength of the qubit-chain coupling in relation to the spectral gap. These are (i) the weak coupling regime, where the qubits are entangled at the asymptotics; (ii) the strong coupling regime, where the concurrence can exhibit collapses followed by revivals with exponentially attenuated amplitude; and (iii) the thermal damping regime, where the concurrence rapidly vanishes due to the chain's thermal excitations. In all cases, if entanglement is generated, this occurs after a finite time has elapsed. This time scale depends exponentially on the qubits distance and is determined by the spectral properties of the chain. Entanglement irreversible decay, on the other hand, is due to the dissipative effect induced by the coupling with the chain and is controlled by the coupling strength between the chain and qubits. This study {unravels the basic mechanisms leading to entanglement in a non-Markovian bath and allows to identify the key} resources for realising quantum coherent dynamics of open systems.  \end{abstract}
\maketitle


\section{Introduction}
\label{secI}
 
The coupling to the surrounding environment is commonly considered the cause of the fragility of quantum superpositions and entanglement. This fragility is a central challenge for quantum technologies \cite{Preskill:2018}, that is {usually} addressed by trying to isolate the quantum system and to actively correct the detrimental effects of the environment on the system's dynamics. This requires an accurate knowledge of the noise induced by the external environment. Within a microscopic theory, the environment is described by a second, large physical system interacting with the system of interest \cite{Gallis:1990,Balian:1991}. These interactions establish entanglement between the system and the environment degrees of freedom, that in turn result in loss of coherence, namely, decoherence \cite{Zurek:2009,Mintert:2005,Konrad:2007}.  Nevertheless, there are counterexamples where the coupling with an external bath can even lead to entanglement between the system's constituents \cite{Plenio:1999,Beige:2000,Braun:2002,Benatti:2003,Cattaneo:2021,Horovitz:2021}. The concepts underlying this environment-induced quantum coherence have been used to design protocols for quantum state preparation and computing using dissipation \cite{Poyatos:1996,Pielawa:2007,Kraus:2008,Verstraete:2009,Menu:2022}. {This progress has} put forward the need to systematically understand what are the features of an environment that are resources, and which are instead detrimental to quantum coherent dynamics. An important resource are decoherence free subspaces, namely, subspaces of the system's Hilbert space that are effectively decoupled from the environment because of destructive interference \cite{Kempe:2001,Lidar2003}. This concept has been recently extended to dynamical symmetries of the master equation, allowing for the existence of stable limit cycles \cite{Buca:2019,Albert:2019}. The backflow of information from environment to the system, characterizing non-Markovianity \cite{Breuer:2016, BRE02}, has been shown to be a further important resource \cite{Rivas:2010,Benatti:2019}. For specific master equations and settings entanglement generation in a non-Markovian environment can be faster than for a Markovian one \cite{Yu:2010,Benatti:2019}. Moreover, non-Markovian baths can distribute entanglement between distant nodes \cite{Plenio:2005,Venuti:2006,Giampaolo:2010,Pielawa:2010,Wolf:2011,Fogarty:2013,Kajari2012,Taketani:2014,Nicacio:2016,Nicacio:2016a,Horovitz:2021}, providing interesting perspectives for realizing quantum communication in non-unitary channels \cite{Bylicka:2014}. In general, classifying the role of the individual features and understanding their interplay would open the perspective to design robust and scalable quantum coherent dynamics in noisy environment. 

\begin{figure}[!htpb]
\includegraphics[width=0.9\columnwidth]{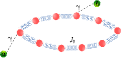}
\caption{(color online) Entanglement generation between two qubits induced by the coupling with a linear chain. The qubits do not mutually interact but are coupled to one oscillator of a linear chain by means of an optomechanical interaction. The chain is a non-Markovian bath that mediates interactions and establishes entanglement between the qubits. The characteristic time scales of entanglement are determined as a function of the spectral gap of the chain, of the strength $\gamma$ of the qubit-oscillator coupling, and of the effective distance $D$ between the qubits, here quantified by the spatial distance between the chain's oscillators to which the qubits couple.}
\label{Fig:1}
\end{figure}

In this work we perform a detailed characterization of a non-Markovian bath in terms of its capability to establish entanglement between two qubits. The non-Markovian enviroment is a chain of oscillators, each qubit couples  to one of the oscillators as illustrated in Fig. \ref{Fig:1} but does not directly couple to the other qubit. Previous studies reported that this configuration may support the onset of entanglement between the qubits \cite{Braun:2002,Plenio:2005,Wendenbaum:2020}, which can even survive at long times \cite{Braun:2001}. The model is amenable to an analytical solution of the propagator \cite{Braun:2001,Wendenbaum:2020}. This permits us to shed light into the individual processes of the bath-induced dynamics, which are otherwise difficult to simulate in a non-Markovian environment \cite{Prior:2010}. In detail, we provide a systematic analysis of the processes that lead to the generation of entanglement and of the ones that cause its decay. We determine the dependence of the corresponding time scales on the strength of the coupling between qubits and chain, on the properties of the chain, and on the distance between the qubits. 

This paper is organised as follows. In Sec. \ref{Sec:2} we introduce the model at the basis of this study and review some of its basic properties. In Sec. \ref{Sec:3} we analyse the dynamics of correlations between the qubits as a function of distance. In Sec. \ref{Sec:4} we characterize the behavior of entanglement as a function of the coupling strength between qubit and chain. We then \gm{determine} the time scales of entanglement generation and decay and their dependence on the physical parameters, including the spatial distance between the qubits. The conclusions are drawn in Sec. \ref{sec:5}. The appendices provide results complementing the discussion in Sec. \ref{Sec:3} and \ref{Sec:4}.


\section{Two qubits coupled to a chain of oscillators}
\label{Sec:2}

The open quantum system is composed by two qubits, here denoted by $a$ and $b$, that do not mutually interact but couple to the vibrational modes of a linear chain. The chain is composed of  $N$ oscillators, the qubits couple to the chain at the oscillators labelled by $\ell_a$ and $\ell_b$, respectively, according to the Hamiltonian:

\begin{equation}
    H_\mathrm{int} = -\hbar \bar\gamma\left(\sigma^a_{x}x_{\ell_a}+\sigma^b_{x}x_{\ell_b}\right)\,, 
\label{eq:H-int}
\end{equation}

\noindent where $\sigma^a_{x}$ and $\sigma^b_{x}$ are the Pauli operators along $x$ for qubits $a$ and $b$ respectively, and $x_\ell$ is the (dimensionless) position operator of the oscillator at site $\ell=1,\ldots,N$ . The parameter $\bar\gamma$ is a positive coupling constant, determining the strength of qubit-chain coupling. The chain dynamics is governed by the Hamiltonian (with periodic boundary conditions)
\begin{eqnarray}
&&H_\mathrm{c} = \frac{\hbar \nu}{2} \sum_\ell  (x_\ell^2 + p_\ell^2) -\frac{\hbar \bar{J}}{2} \sum_\ell x_\ell x_{\ell+1}\,, 
\label{H-pos}
\end{eqnarray}
where $p_\ell$ is the dimensionless canonically conjugated operators to $x_\ell$, while the positive frequencies $\nu$ and $ \bar{J}$ respectively scale the energy of the oscillators and  the coupling between neighboring oscillators. The chain's stability is warranted by the inequality $ \bar{J}\le \nu$. The ratio $\bar J/\nu$, in particular, controls whether the chain's spectrum is gapped ($ \bar{J}<\nu$) or gapless ($ \bar{J}=\nu$). The system's Hamiltonian is here neglected and the total Hamiltonian is thus given by $H=H_\mathrm{c}+H_\mathrm{int}$: The Hamiltonian commutes with the operators $\sigma^j_x$ and the chain hence acts as a (non-Markovian) dephasing bath in the eigenbasis of the operators $\sigma^j_x$. 

In the following, we rescale the energy by $\hbar \nu$ and the time by $\nu^{-1}$. The rescaled oscillator-oscillator and qubit-oscillator couplings are  $J_0=\bar{J}/\nu$ and $\gamma=\bar{\gamma}/(\sqrt{2}\nu)$, respectively. We further introduce the distance $D$ between the qubits in terms of the distance between the oscillators to which they couple,  $$D=|\ell_b-\ell_a|\,.$$ This quantity corresponds to the number of oscillators separating oscillator $\ell_a$ from oscillator $\ell_b$. For later convenience, we will use the definition $\ell_a=N-l+1$ and $\ell_b=l$, with $l=1,\ldots,\frac{N}{2}$ for $N$ even ($l=1,\ldots,\frac{N+1}{2}$ for $N$ odd). According to this notation, the distance between the qubits can be rewritten as $D=N+1-2l$.


\subsection{Symmetries}

In order to determine the propagator, we make use of the symmetry of the total Hamiltonian by reflection about the centers of the chain segments that connects the two qubits. Given the periodic boundary conditions, there are two such points and we choose the one separating the shortest segment. This symmetry defines the separated subspaces containing respectively the symmetric and antisymmetric states under this reflection. In order to single out this property we introduce the chain's symmetric and antisymmetric coordinates
\begin{eqnarray}
&&x_{N-2\ell +1}^{S} = \frac{1}{\sqrt{2}}\left(x_{N -\ell+1} +x_\ell \right)\,, \\
&&x_{N-2\ell +1}^{A} = \frac{1}{\sqrt{2}}\left(x_{N -\ell +1} -x_\ell \right)\,,
\end{eqnarray}
and the respective canonically-conjugated momenta, $p_{N-2\ell +1}^S$ and $p_{N-2\ell +1}^A$. The chain Hamiltonian can be decomposed into the sum of the Hamiltonian for the symmetric and the antisymmetric modes, $H_\mathrm{c}=H_\mathrm{c}^S+H_\mathrm{c}^A$, which mutually commute: $[H_\mathrm{c}^S,H_\mathrm{c}^A]=0$. Using these coordinates, the interaction Hamiltonian, Eq.\ \eqref{eq:H-int}, takes the form
\begin{equation}
H_\mathrm{int} ^{(D)}= -\sqrt{2} \gamma ( \sigma_x^a+\sigma_x^b)x_D^S-\sqrt{2} \gamma ( \sigma_x^a-\sigma_x^b) x^A_D \,,
\label{H-int-symI}
\end{equation}
for $D>0$. For the given Hamiltonian, the operators $\sigma_x^{a,b}$ are constants of motion. Therefore, a convenient qubits basis for analysing the dynamics is the basis of eigenstates 
$\ket{ij}\equiv \ket{i}_a\otimes\ket{j}_b$, with $\sigma_x^{a,b}\ket{\pm}_{a,b}=\pm\ket{\pm}_{a,b}$. In particular, the two-dimensional subspace spanned by the basis vectors $ \{\ket{b^S} \}: \{\ket{++}_x, \ket{--}_x\} $ couples to the symmetric chain, while the antisymmetric chain couples to the two-dimensional subspace spanned by the basis vectors $\{\ket{b^A}\}:\{\ket{+-}_x, \ket{-+}_x\} $. In this basis the coupling does not modify the occupation of the eigenstates, but affect the evolution of superpositions. The states $ \{\ket{b^S} \}$ are eigenstates of the operator $S_x^S=(\sigma_x^a+\sigma_x^b)/2$ at the eigenvalues $b^S=\pm 1$ and belong to the kernel of operator $S_x^A=(\sigma_x^a-\sigma_x^b)/2$. Vice versa, the states $ \{\ket{b^A} \}$ are eigenstates of the operator $S_x^A$ at the eigenvalues $b^S=\pm 1$ and belong to the kernel of operator $S_x^S$. This shows that in general no decoherence-free subspace exists for $D>0$. The case $D=0$ is special. In this case
\begin{equation}
H_\mathrm{int} ^{(0)}= -\gamma  (\sigma_x^a+\sigma_x^b)x_0\,,
\label{H-int-symII}
\end{equation}
and there is a decoherence-free subspace of the qubits Hilbert space, which consists of the kernel of operator $\sigma_x^a+\sigma_x^b$. 


\subsection{Chain's normal modes}

Before proceeding, it is useful to introduce the normal mode coordinates $\tilde{x}_n^S$,  $\tilde{p}_n^S$ and  $\tilde{x}_n^A$,  $\tilde{p}_n^A$, with normal mode frequency $\omega_{n}^{j=S,A}=\sqrt{1-J_0\cos k_n^j}$ and wave numbers $k_n\in [-\pi,\pi)$. As we will consider finite chains, one shall distinguish between an even or odd numbers of oscillators. In detail: for $N$ even, then the Brillouin zones for the symmetric and antisymmetric modes are $k_n^S = 2n\pi/N$ and $k_n^A =k_n^S + 2\pi/N$ with $n=0,\ldots, \frac{N}{2} - 1$. For $N$ odd, then $n=0,\ldots, \frac{N-1}{2}$ for $k_n^S$ and  $n=0,\ldots, \frac{N-3}{2}$ for $k_n^A$.

In the normal mode representation, the symmetric and antisymmetric part of the chain's Hamiltonian read
\begin{eqnarray}
\tilde{H}^{j=S,A} &=& \frac{1}{2} \sum_{n} \left(  \tilde{p}_{n}^{j2} + \omega_{n}^{j2}\tilde{x}_{n}^{j2}\right)\,.
\end{eqnarray}

We write the interaction Hamiltonian $H_i$ using the chain's normal modes: 
\begin{equation}
    H_\mathrm{int}=-\sum_{n}\left(\tilde{\gamma}_n^{S} \tilde{x}_n^S ( \sigma_x^a+\sigma_x^b)/2+\tilde{\gamma}_n^{A} \tilde{x}_n^A( \sigma_x^a-\sigma_x^b)/2 \right)\,.
    \label{eq:H_int-normal}
\end{equation}

The coupling constants now depend on the normal modes and read:
\begin{eqnarray}
\label{eq:gamma:D}
&&\tilde{\gamma}^S_n=\frac{2 \sqrt{2}\gamma}{\sqrt{N}}\cos\left(\frac{k_n^SD}{2}\right)\,, \nonumber\\
&&\tilde{\gamma}^A_n=\frac{2 \sqrt{2}\gamma}{\sqrt{N}}\sin\left(\frac{k_n^AD}{2}\right)\,.
\end{eqnarray}
The coupling to the mode at lowest frequency ($n=0$) is $\tilde{\gamma}_0^S = \frac{2 \gamma}{\sqrt{N}}$, and the coupling to the mode at largest frequency ($n=N/2$) takes the form $\tilde{\gamma}^A_{N/2}|_{\text{even}} = \frac{2 \gamma}{\sqrt{N}} (-1)^{\frac{D+1}{2}}$ and $\tilde{\gamma}^A_{(N-1)/2}|_{\text{odd}} = \frac{2 \gamma}{\sqrt{N}} (-1)^{\frac{D}{2}}$ (the latter distinction is a finite-size effect). Interestingly, the distance introduces a characteristic wave number $k_D=2\pi/D$ which modulates the coupling strength to a given mode.  

We will analyse the dynamics as a function of the coupling strength $\gamma$, of the distance $D$, and of the elastic strength $J_0$. The latter controls the gap of the chain's spectrum, $$\omega_0=\sqrt{1-J_0}\,.$$ 
In this work we choose $J_0<1$, thus the spectrum is gapped. The coupling $J_0$ also determines the bandwidth $\Delta\omega=\sqrt{1+J_0}-\sqrt{1-J_0}$, as well as the velocity with which information propagates along the chain. For $\gamma=0$ this is quantified by the Lieb-Robinson bound, which gives the maximal velocity $v_{\text{LR}}$ with  which information can propagate through a non-relativistic quantum system \cite{LiebRobinson}:
\begin{equation}
\label{eq:LR}
v_{\text{LR}} =  2 \max_{k \in BZ} | \nabla_k \, \omega_k|,
\end{equation}
where $BZ$ here indicates the Brillouin zone. Note that $v_{\rm LR}$ vanishes for $J_0\to 0$ and monotonously increases in the interval $J_0\in [0,1]$.


\subsection{Propagator}
\label{secIII}

Let $\rho_d(t)$ be the density matrix of the two qubits at time $t$. It is obtained by tracing out the chain's degrees of freedom from the density matrix $\chi(t)$ of the total system, $\rho_d(t) = \text{Tr}_c [\chi(t)]$, where $ \text{Tr}_c$ denotes the trace over the chain's degrees of freedom. In this work we assume that at $t=0$ the initial density matrix $\chi(0)$ is a separable state of qubits and chain, $\chi(0)=\rho_d(0)\otimes R_\beta$. The propagator $\Lambda_t$, connecting the state of the qubits at time $t$ with the initial state $\rho_d(0)$, is defined as
\begin{equation}
\label{rho:d}
\rho_d(t) = \text{Tr}_c [{U}(t)\rho_d(0)\otimes R_\beta{U}^{\dagger}(t)]\equiv\Lambda_t\rho_d(0)\,.
\end{equation}
Here, $U(t)=\exp(-{\rm i}(H_\mathrm{int}+H_\mathrm{c})t/\hbar)$ is the evolution operator. An analytical form for the propagator is found when the chain is in a Gaussian state. In what follows the chain is initially prepared in a thermal state with inverse temperature $\beta$, namely, $R_\beta=\exp(-\beta H_c)/Z$, with $Z$ being the partition function. The propagator is conveniently expressed in the basis of eigenstates $\{\ket{b_i}\} : \{\ket{++}, \ket{--}, \ket{+-}, \ket{-+}\}$. Recalling that the symmetric (antisymmetric) subspace is $ \{\ket{b^S} \}: \{\ket{++}_x, \ket{--}_x\} $ (respectively $\{\ket{b^A}\}:\{\ket{+-}_x, \ket{-+}_x\} $), the matrix elements coupling states in the same subspace evolve according to \cite{Braun:2001,Wendenbaum:2020}
	\begin{eqnarray}
		&&\bra{b_i^{\alpha}}\rho_{d}(t)\ket{b_j^{\alpha}}  = \bra{b_i^{\alpha}}\rho_{d}(0)\ket{b_j^{\alpha}}    \label{eq:evolution} \\
		&& \times  \exp\left(-f^{\alpha}(t)\left(b_i^{\alpha}-b_j^{\alpha}\right)^2\right)\,,\nonumber
	\end{eqnarray}
where $\alpha = S,A$ and
\begin{align}
f^\alpha(t) &=\frac{1}{2} \sum_n \frac{\tilde{\gamma}_n^{\alpha\,2}}{\omega_n^{\alpha\,3}}\left(2\bar n(\omega_n^\alpha) + 1\right)\left(1-\cos\left(\omega_n^\alpha t \right)\right) \,, \label{eq:fsa}
\end{align}
with the mean thermal occupation of the $\alpha$-th symmetric (antisymmetric) mode of the chain being given by the Bose-Einstein statistics $\bar n(\omega)  = 1 / (e^{\beta\omega} - 1 )$. The dynamics leads to damping of the off-diagonal elements within the same subspace, while the diagonal elements are constant in time. We will denote this term by {\it damping term} (or also {\it attenuation}).

The time evolution of the matrix elements between the symmetric and the antisymmetric subspaces takes instead the form:
\begin{eqnarray}
&&\bra{b_i^{\alpha}}\rho_{d}(t)\ket{b_j^{\beta}}_{\beta\neq\alpha}  = \bra{b_i^{\alpha}}\rho_{d}(0)\ket{b_j^{\beta}}_{\beta\neq\alpha}   \label{eq:evolution:off} \\
&& \times  \exp\left(-f^{\alpha}(t)-f^{\beta}(t)+i(\varphi^{\alpha}(t)-\varphi^{\beta}(t))\right)\,, \nonumber
\end{eqnarray}
where the time-dependent phases read  \cite{Braun:2001,Wendenbaum:2020}
\begin{align}
%
\varphi^\alpha(t) &= \frac{1}{2}\sum_n\left(\frac{\tilde{\gamma}_n^{\alpha\,2}}{\omega_n^{\alpha\,2}}t-\frac{\tilde{\gamma}_n^{\alpha\,2}}{\omega_n^{\alpha\,3}}\sin\left(\omega_n^{\alpha}t\right)\, \right)\,.\label{eq:phisa}
\end{align}
The phase is the sum of two contributions: (i) a contribution linear in time and (ii) a multi-chromatic, oscillating contribution. The term (i) gives rise to a periodic oscillation at frequency $\Omega_0=\Omega_0^S-\Omega_0^A$ with 
\begin{equation}
\Omega_0	^\alpha=\sum_n\frac{\tilde{\gamma}_n^{\alpha\,2}}{\omega_n^{\alpha\,2}}\,.
\end{equation}
The oscillations can be associated to an effective Hamiltonian dynamics similarly to the Lamb-shift Hamiltonian of quantum electrodynamics \cite{Bethe:1947,Buchheit:2016}. Due to this analogy and to its collective nature (see the following section), we will denote this frequency by {\it collective Lamb shift}  \cite{Gross:1982,Scully:2009,Konovalov:2020}. For a single qubit the multichromatic oscillations are responsible for decoherence \cite{Braun:2001}. For this reason we will refer to this term as {\it decoherence} or also {\it dephasing term}.


\subsection{Discussion}

The form of the propagator allows us to identify some relevant time scales. We first observe that real and imaginary parts exhibit multichromatic oscillations at the frequency of the chain's spectrum. At the time $t\ll t_{\rm max}\equiv 1/\omega_{\rm max}$ one finds the scaling behavior $f\propto t^2$ while $\varphi\propto t^3$, which is independent of the chain spectrum \cite{Strunz:2003,Braun:2001}. During the evolution, as time goes by the influence of the chain's normal modes at decreasing frequency start to be important. When $\Delta\omega\gg\omega_0$, one can identify a time scale separation at which the low-frequency modes of the reservoir spectrum become relevant in determining the system evolution \cite{Gualdi:2013}.  In this work this is not the case, since the chain spectrum is gapped, $\omega_0>0$ and the largest frequency of the normal modes, $\omega_{\rm max}=\omega_0+\Delta\omega$, is chosen to be of the same order as $\omega_0$. For our dynamics, an important time scale is $t_{\rm chain}\sim \pi/\Delta \omega$, at which the oscillations at the smallest and at the largest chain frequencies are out of phase by $\pi$. At this time scale the decoherence and damping terms start to oscillate.  There is a further important time scale to consider in our analysis. This is the time scale over which the finite size of the chain becomes important and is essentially the  Poincar\'e time \cite{Balian:1991}. We estimate the Poincar\'e time using the Lieb-Robinson bound, $t_P\sim N/(2v_{\rm LR})$. For $N\gg 1$ and times $t\ll t_P$ the dynamics is independent of the specific chain's size. 

Let us now discuss the real and imaginary parts of the propagator  separately. The dissipative component scales with $\bar n(\omega)$ and thus increases with the initial temperature of the chain. Its value as a function of time is bound from above: $f\le 16(2\bar n(\omega_0)+1)\gamma^2/\omega_0^3$. Hence, for weak couplings and moderate temperatures  the off-diagonal elements of the density matrix are attenuated with respect to the initial value, while for large couplings they vanish after a time $t\gg t_{\rm max}$. The function $\varphi$, instead, is independent of the temperature. Interestingly, for a single qubit it vanishes identically \cite{Braun:2001}. For two qubits, this term vanishes when the qubits are either in an eigenstate $|ij\rangle$ of $\sigma_x^{a,b}$ or in a maximally entangled state. This also implies that the frequency $\Omega_0$ is different from zero only in the presence of a second qubit. Note that the frequency $\Omega_0$ corresponds to an effective, coherent interaction between the qubits that emerges because of the coupling with the bath. 


\section{Bath-induced dynamics}
\label{Sec:3}

The physical problem here considered is an example of bath-induced dynamics, where the reservoir is intrinsically non-Markovian. Notably, for a single qubit the dynamics is solely dissipative, while coherent oscillations and decoherence are induced only in the presence of a second qubit. In order to identify the salient physical regimes, it is useful to first consider three asymptotic, well-defined limits, taking care that the symmetry of the two-qubit dynamics is preserved. (i) In the trivial limit $\gamma= 0$ the qubits are isolated  and undergo no dynamics. (ii) When $\overline{J}\to 0$ and at finite $\gamma$, the oscillators of the chain become decoupled, and the dynamics reduces to a qubit coupled to a single oscillator at frequency $\omega_0=1$. In this case, correlations can be established between the qubits only when they couple to the same oscillator, namely, for $D=0$. (iii) Finally, in the limit $\gamma\to\infty$ the dynamics reduces to an exchange of excitations between qubit and the oscillator to which it couples. The interaction Hamiltonian tends to freeze the oscillator in an eigenstate of the position operator and to suppress propagation of excitations along the chain. Outside of these asymptotic cases, the dynamics is characterized by oscillations at frequency $\Omega_0$ that are dephased by the second term in the function $\varphi$ and damped by the attenuation function. 


\subsection{Weak and strong coupling regimes}

Some salient regimes can be identified for a gapped chain spectrum ($\overline{J}<1$), which is the situation we analyse in this work. For $\gamma\ll\omega_0^{3/2}$, the coupling to the chain gives rise to the coherent interaction $\Omega_0$ while decoherence and damping (at moderate temperatures) are small corrections that limit the visibility of the oscillations. We will denote this regime by weak-coupling regime.  In the opposite limit, the strong coupling regime $\gamma\gg\omega_0^{3/2}$, damping and decoherence become dominant at sufficiently long times, which we identify with $t\gg t_{\rm max}$. There is a third regime, where the temperature is such that $\gamma\ll\omega_0^{3/2}$ but $\gamma \sqrt{\bar n} \gtrsim \omega_0^{3/2}$: in this case one observes decaying oscillations due to damping, while dephasing is negligible. The oscillations vanish at the asymptotics when $\gamma \sqrt{\bar n} \gg \omega_0^{3/2}$. We will denote this regime by thermal-damping regime. 
These considerations are useful for understanding the dynamics of observables and correlations. 


\subsection{Dynamics of the correlation functions}

{For a single qubit the coupling with the chain induces decoherence and dissipation, but no dynamics \cite{Strunz:2003}. Interestingly, the coupling of a second qubit gives rise to observable effects on the first qubit. In this section we discuss these dynamics by analysing} the single-particle expectation values $\langle \sigma_\alpha^{a,b}\rangle$ and the two-particle correlation function,
\begin{equation}
\label{Eq:g2}
	g_{\alpha,\beta}(t,t')=\langle \sigma_\alpha^a(t) \sigma_\beta^b (t')\rangle - \langle \sigma_\alpha^a(t)\rangle \langle\sigma_\beta^b (t')\rangle\,,
\end{equation}
where $\langle \cdot\rangle={\rm Tr}\{\cdot \rho_d(0)\}$. The qubits are initially aligned along $y$, namely, they are prepared in the pure states $\ket{\psi}_{j=a,b}=(\ket{+}_x+\ket{-}_x)/\sqrt{2}$. Moreover, we assume that the chain is at low temperatures, such that the mean occupations of the normal modes are $\bar n(\omega)\ll 1$. 

\begin{figure}
	\includegraphics[scale= 0.47]{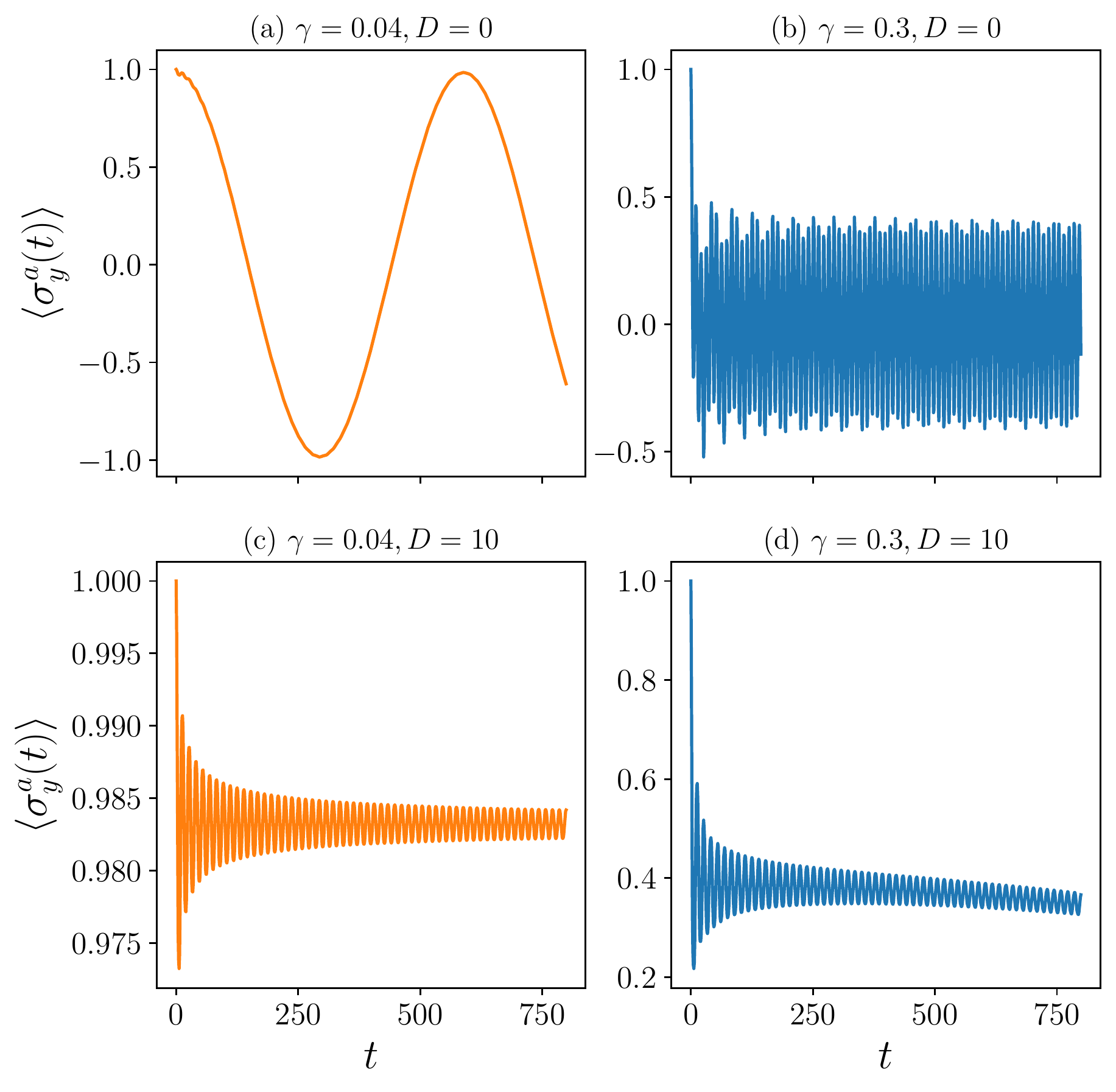}
	\caption{Dynamics of the observable $\langle \sigma_y^a(t) \rangle $ when the qubits are at distance $D=0$ (upper row) and at distance $D=10$ (lower row). The coupling strength is $\gamma=0.04$ in the left subplots, (a) and (c),  and $\gamma = 0.3$ in the right subplots (b) and (d). The chain consists of $N = 10^4$ spins with $\omega_0 = 0.45$ and inverse temperature $\beta = 10^5$. The qubits are initially prepared in the pure state $\ket{\psi}=\ket{\psi}_a\otimes \ket{\psi}_b$ with $\ket{\psi}_{j=a,b}=(\ket{+}_x+\ket{-}_x)/\sqrt{2}$, corresponding to eigenstates of $\sigma_y^{a,b}$ at eigenvalue 1.}
		\label{fig:corr1}
\end{figure}

Figure \ref{fig:corr1} shows the dynamics of the $y$-component of one of the qubits in the weak (left) and in the strong (right) coupling regime, when the second qubit is at distance $D=0$ (upper row) and $D>0$ (lower row). We first discuss the case $D=0$, where there is a decoherence-free subspace. In the weak coupling regime, Fig.\ \ref{fig:corr1}(a), the qubit undergoes slow oscillations at the frequency $\Omega_0$, whose visibility is close to unity. {We emphasize that the qubit oscillation is a cooperative effect determined by the collective Lamb shift.} 

The strong coupling regime is shown in Fig.\ \ref{fig:corr1}(b): The oscillations become multichromatic, {the characteristic frequencies are the normal modes of the chain which are excited during the corresponding time scale, the corresponding} amplitude is quickly damped to the asymptotic value. 

When instead the qubits are at {finite} distance, {here} $D=10$, the dynamics at weak coupling is effectively frozen out, see Fig.\ \ref{fig:corr1}(c), while at strong coupling the expectation value first quickly decays to a non-vanishing value, and then undergoes a slow dynamics. This behavior for relatively large distances is understood by inspecting the dependence on $D$ of the parameter $\bar \gamma_n$, Eq. \eqref{eq:gamma:D}, which scales the Fourier components of the propagator. The damping of the symmetric and antisymmetric modes is multiplied by the factor ${\cos^2(kD/2)}$ and ${\sin^2(kD/2)}$, respectively, while the Fourier components of the imaginary part, $\varphi^S-\varphi^A$, are scaled by the factor ${\cos(kD)}$. In particular, 
\begin{equation}
\label{Eq:Omega:D}
\Omega_0(D)\simeq 4\gamma^2/N\sum_n\cos(k_nD)/\omega_n^2\,.
\end{equation} 
As $D$ increases the value of $\Omega_0$ decreases and for the spectrum $\omega_n$ here considered, then $\Omega_0\to 0$ for $D\to\infty$. 
Therefore, for $D\gg 1$ the oscillation frequency is such that $|\Omega_0|\ll 1$, and in the weak coupling limit the dynamics is essentially frozen over the considered time scale. In the strong coupling limit, Fig.\ \ref{fig:corr1}(d), the expectation value quickly decreases due to damping, while the slower dynamics is due to the slow oscillation at frequency $\Omega_0$.

\begin{figure}
	\includegraphics[scale= 0.47]{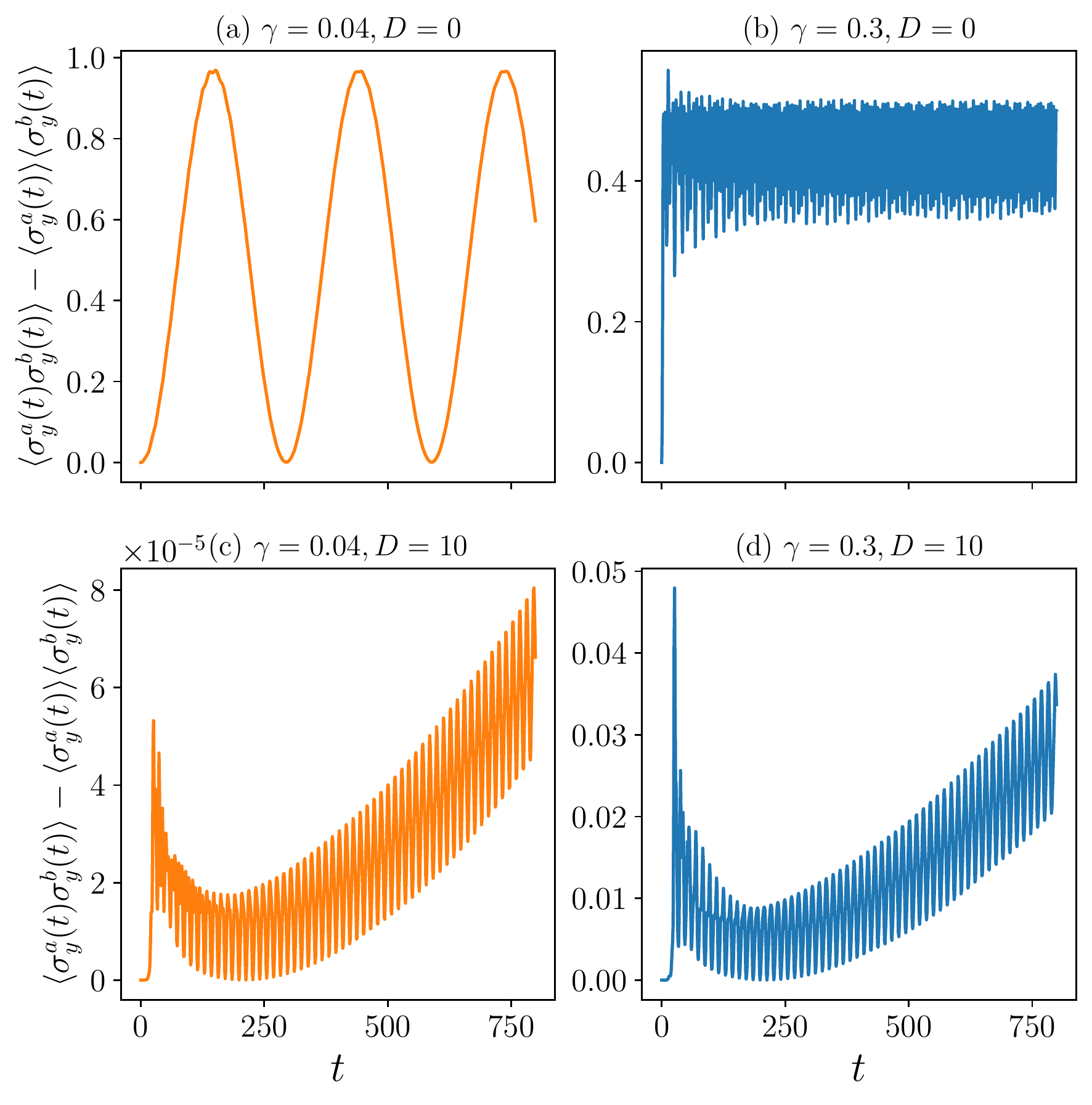}
	\caption{ Qubit-qubit correlations, \eqref{Eq:g2} for $\alpha=\beta=y$ and at equal time, $g_{yy}(t,t)$. The parameters of the subplots are the same as in Fig. \ref{fig:corr1}.}
	\label{fig:corr2}
\end{figure}

Figure \ref{fig:corr2} displays the correlations between the two qubits, corresponding to the subplots of Fig. \ref{fig:corr1}. The correlations are here evaluated at equal time, and we choose to show the representative case $\alpha=\beta=y$. Starting from a separable state, we observe that the correlations grow with $t^3$ at very short times. This demonstrates that correlations are first established by the imaginary component of the propagator. After this transient, in subplots (a), (c) and (d) they exhibit a slow oscillation at the characteristic frequency $\Omega_0$, about which we observe fast oscillations at the normal mode frequencies of the chain (which are not visible in the weak coupling regime, subplot (a), because the amplitude is very small). The strong coupling regime for $D=0$, Fig.\ \ref{fig:corr2}(b), shall be discussed apart. In this case damping and dephasing are dominant and the correlations quickly reach a stationary state about which fast oscillations occur. We refer the reader to Appendix \ref{App:A} for the analytic expression of all two-point correlations at equal and different times. {These results allow us to identify mechanisms, which determine the dynamics of entanglement. Interestingly, they also show that the measurement of single and two-body correlations provide an insightful probe of the surrounding, non-Markovian environment. }

\section{Entanglement}
\label{Sec:4}

Having shown that the chain establishes correlations between the qubits, we verify when these correlations are non-classical. {Several measures have been discussed in the literature for studying entanglement between  qubits \cite{Horodecki:2009}. For our purpose, the concurrence is a useful quantity, since for two qubits it also quantifies entanglement. The concurrence is analysed as a function of time and is defined as}
 \cite{Wooters1998}: 
\begin{eqnarray}
C(\rho(t)) = \max\{0, \lambda_1 - \lambda_2 - \lambda_3 - \lambda_4\},
\end{eqnarray}
where $\lambda_1, \lambda_2, \lambda_3, \lambda_4$ are the eigenvalues, in decreasing order, of the matrix $\omega(t)=\sqrt{\sqrt{\rho_d(t)} \tilde{\rho}_d(t)\sqrt{\rho_d(t)}}$ with $\tilde{\rho}_d(t) = (\sigma_y \otimes \sigma_y)\rho_d^{*}(t) (\sigma_y \otimes \sigma_y)$ and the complex conjugate is taken in the eigenbasis of the Pauli matrix $\sigma_z$. The concurrence being a monotonous function of entanglement, it provides information about the growth or decay of quantum correlations between the qubits. The definition we use, moreover, applies also to mixed states. 

In order to analyse the capability of the chain to mediate entanglement, we assume that the qubits are initially uncorrelated. Their initial density matrix is $\rho_d(0)=\rho_a(0)\otimes \rho_b(0)$, with $\rho_{j=a,b}(0)$ the density matrix of qubit $a,b$ and $\rho_{j}(0)=\ket{\Psi}_j\bra{\Psi}$ is a pure state. We parametrize the initial state in the Bloch-sphere representation
 \begin{equation}
\ket{\Psi}_j=\cos\theta_{j}\ket{+}_x+e^{i\phi_{j}}\sin\theta_{j}\ket{-}_x, \label{eq_bloch}    
\end{equation}
with $\theta_j,\phi_j$ real and $\ket{+}_x$ and $\ket{-}_x$ the eigenvectors of $\sigma_x$ Pauli matrix with eigenvalues $\pm 1$ respectively. 

We first focus on the characteristic time scales of entanglement for the case in which there exists a decoherence-free subspace for the qubits, namely, the qubits couple to the same oscillator. We then analyse the characteristic length scales by studying the situation where the qubits couple to different oscillators as a function as the distance $D$ between the oscillators.


\subsection{Time scales of entanglement}

When the qubits couple to the same oscillator, $D=0$, the antisymmetric subspace is decoupled from the chain and is a decoherence-free subspace. Deep into the weak-coupling regime, in leading order the dynamics is coherent and the concurrence is given by $C(\rho) = \sqrt{2(1 - \text{Tr}\{ \rho_R^2 \})}$ where $\rho_R$ is the reduced density matrix of one qubit obtained after tracing out the Hilbert space of the other one \cite{Horodecki:2009}. When the qubits are initially prepared in the initial state \eqref{eq_bloch}, then the concurrence takes the form 
\begin{equation}
\label{Eq:C:analytical}
	C(\rho(t)) \simeq |\sin(2 \theta_a) \sin(2 \theta_b) \sin(\Omega_0 t) |\,,
\end{equation}
and periodically oscillates at the frequency of the collective Lamb shift $\Omega_0$. The periodic oscillation characterizes also the dynamics of the correlations in the weak-coupling regime and indicates that the collective Lamb-shift  term of the propagator is responsible for the appearance of entanglement. The initial state of the qubits determines the maximum reached by the concurrence: The concurrence varies between 0 and $C_{\rm max}= |\sin(2 \theta_a) \sin(2 \theta_b) |$ with period $\pi/\Omega_0$ and reaches the maximal amplitude $C_{\rm max}=1$ for initial states with Bloch angles $\theta_a,\theta_b=(2n+1)\pi/4$, with $n\in \mathbb{Z}$, namely, when the qubits are aligned along $y$ or $z$. 

We note that Eq.\ \eqref{Eq:C:analytical} is valid after a transient time, and specifically for $t\gtrsim t_{\rm max}$. For $t\ll t_{\rm max}$, instead, the term $\Omega_0t$ is of the same order as the dephasing term. This is the regime where decoherence exhibits the universal behavior $\varphi\propto t^3$ and where we do not expect to find entanglement because of dephasing. This argument shows that there is a finite time scale at which entanglement is generated between the qubits. We will extensively characterize it in the next subsection. 

\begin{figure*}[!htpb]
  \includegraphics[width=\textwidth,height=5cm]{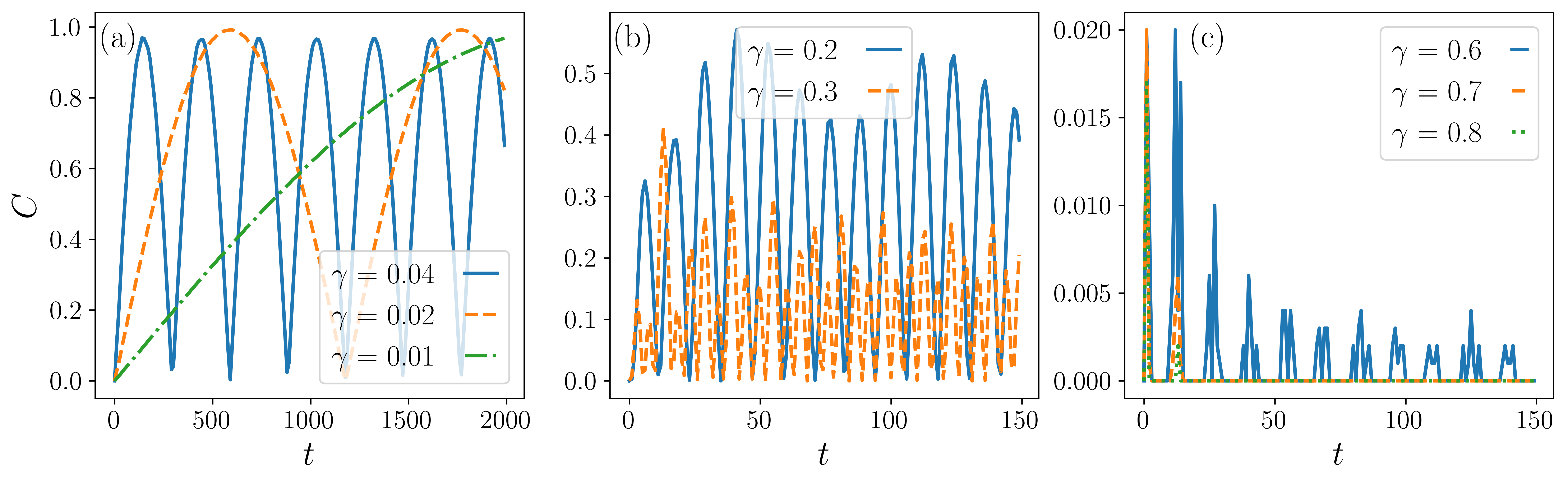}
  \caption{Concurrence $C$ for different values of $\gamma$ when the qubits are coupled to the same oscillator, $D=0$. Subplot (a) shows $C$ for three different values of $\gamma$ deep in the weak coupling regime, subplot (b) shows the concurrence for $\gamma\simeq\omega_0^{3/2}$, where the effects of damping and decoherence starts becoming relevant on the time scale of the oscillations, and subplot (c) displays the concurrence for $\gamma\gg\omega_0^{3/2}$, in the regime dominated by damping and decoherence. The qubits are initially prepared in the same pure state \eqref{eq_bloch} with $\theta_a = \theta_b = \pi/4$ and $\phi_a = \phi_b = 0$. The chain is composed by $N=10^4$ oscillators with $\omega_0=0.45$ and inverse temperature $\beta = 10^5$.}
\label{fig:concurrence}
\end{figure*}

We now discuss the behavior of the concurrence for different values of the coupling strength $\gamma$, ranging from the weak- to the strong-coupling regime. In what follows we assume that the qubits are initially prepared in the same eigenstate of $\sigma_y$. Unless otherwise stated, the chain is initially at very low temperatures.  Figure \ref{fig:concurrence}(a) displays the time evolution of the concurrence deep in the weak coupling regime, for $\gamma\ll\omega_0^{3/2}$.  The concurrence displays a periodic oscillation as in Eq. \eqref{Eq:C:analytical}, with slightly reduced visibility due to damping. There are also fast oscillations about the slow envelope, that become visible only after zooming in and are due to the dephasing term of the propagator.  For $\gamma\sim\omega_0^{3/2}$, Fig.\ \ref{fig:concurrence}(b), the concurrence exhibits a multichromatic oscillation whose mean amplitude decreases with $\gamma$. At even larger values, deep in the strong coupling regime,  the concurrence features collapses and revivals with quickly decreasing amplitude, see Fig.\ \ref{fig:concurrence}(c). The decay of the signal is due to the attenuation, the collapses and revivals are due to dephasing and rephasing of the function $\varphi$. The rephasing time is mostly determined by the beating of $\Omega_0$ and $\omega_0$. The dephasing time scales with $t_{\rm chain}$ and is determined by the finite bandwidth $\Delta\omega$ of the chain's spectrum.  

In order to study the {entanglement} at long times we use the time average concurrence $\bar C$, which we define as
\begin{equation}
\label{Eq:bar:C}
\bar C=\frac{1}{t_{\rm end}}\int_0^{t_{\rm end}}C(t')dt' ,
\end{equation}
where $t_{\rm end}$ is the integration time. In order to capture a sufficiently large number of oscillations, {the dynamics is evolved over} times $t_{\rm end}$ which are generally longer than the Poincar\'e time $t_P$. We show in Appendix \ref{App:B} that the behavior found at $t_{\rm end}$ is in qualitative agreement with the one found for $t<t_P$. The results we are going to discuss, hence, give a reliable indication of the be havior in the thermodynamic limit {at finite $\gamma$. The behavior in the region at $\gamma\to 0$, instead, is determined by finite-size effects, as we will argue below.} 

\begin{figure}
	\includegraphics[scale =0.5]{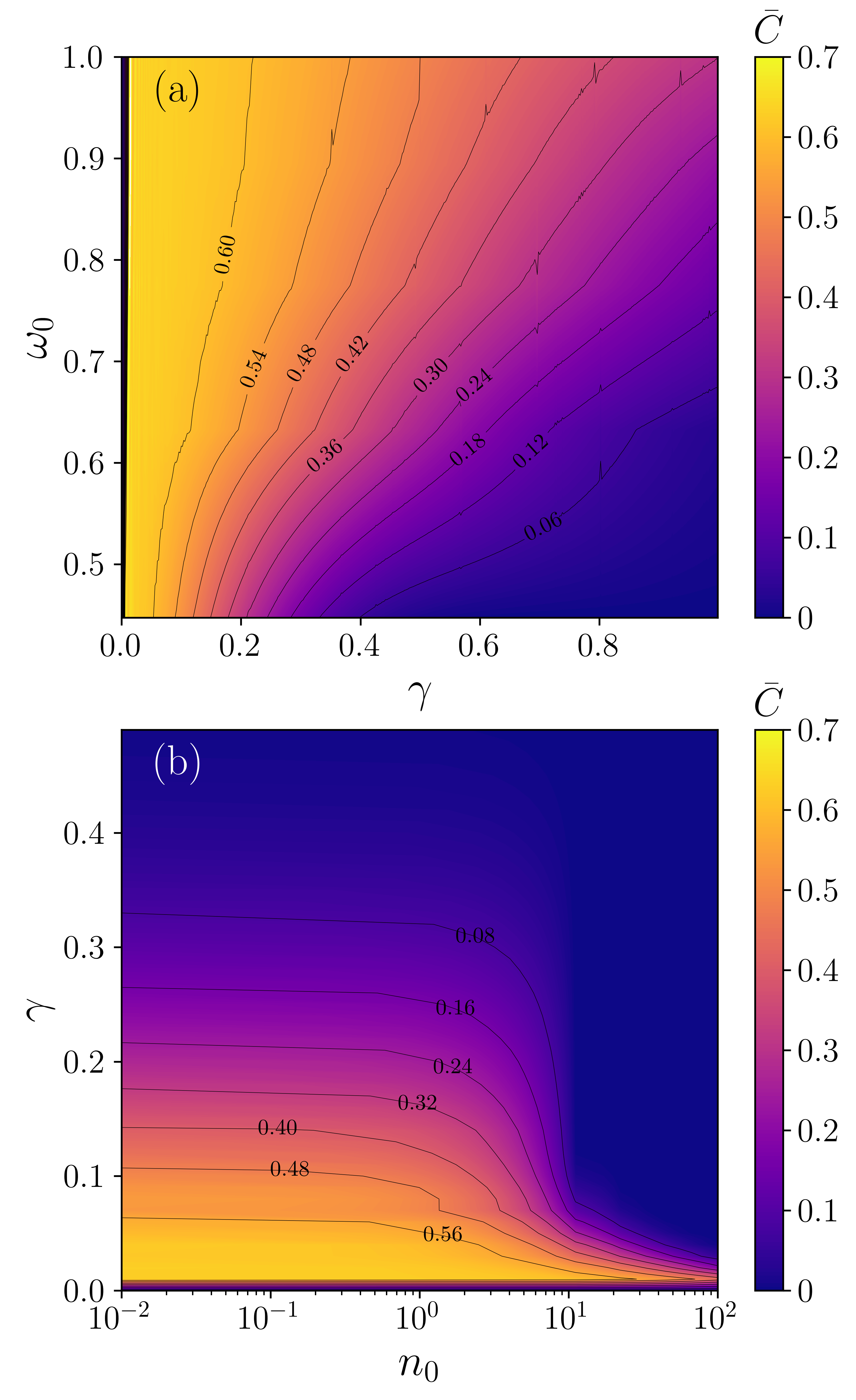}
	\caption{(color online) Color plot of the average concurrence $\bar C$, Eq. \eqref{Eq:bar:C} for $D=0$. Subplot (a) shows $\bar C$ as a function of $\gamma$ and $\omega_0$ for $\beta=10^5$. Subplot (b) displays $\bar C$ as a function of $\gamma$ and $n_0\equiv n(\omega_0)$, the average occupation of the mode at frequency $\omega_0$, for $\omega_0=0.45$. The plots have been evaluated for a chain with $N=10^4$ oscillators and integrating over the time $t_{\rm end}= 10^6$. The qubits are initially in the same initial state with Bloch angles $\theta_a = \theta_b = \pi/4$ and $\phi_a = \phi_b = 0$.}
	\label{fig:phase_diag_same}
\end{figure}

Figure \ref{fig:phase_diag_same}(a) displays the entanglement phase diagram as a function of the coupling strength $\gamma$ and of the characteristic frequency $\omega_0$ of the chain. The diagram qualitatively correspond to the steady state ($t_{\rm end}\to\infty$) with the exception of the small stripe at $\gamma\to 0$. In this region the average concurrence vanishes because the time scale, at which entanglement is generated, is larger than the integration time $t_{\rm end}$ (In  Appendix \ref{App:B} we show that the size of this region shrinks as $t_{\rm end}$ increases). Outside of this region, the concurrence is maximum deep in the weak coupling regime, while it decays as $\gamma$ is increased. More specifically, the contour lines correspond to good approximation to the constant values $\gamma^2/\omega_0^3$: The average concurrence monotonically decreases to zero as the ratio $\gamma^2/\omega_0^3$ grows, and with it the role of dissipation and dephasing on the dynamics. 

Figure \ref{fig:phase_diag_same}(b) shows the effect of the initial chain temperature, here given by the mean occupation $n_0\equiv n(\omega_0)$, on the asymptotic behavior of entanglement.  At low temperatures, $n_0<1$, the average concurrence is almost independent of the temperature: it decays to zero as $\gamma$ reaches the strong coupling regime. By increasing $n_0$ above unity the system enters the thermal damping regime: entanglement vanishes at lower values of $\gamma$. The contour lines, in particular, follow the functional behavior $\gamma\sqrt{n_0}=$ constant, and show that damping is majorly responsible for the disappearance of the average concurrence. 

\subsection{Entanglement as a function of the distance}

\begin{figure}
\includegraphics[scale= 0.5]{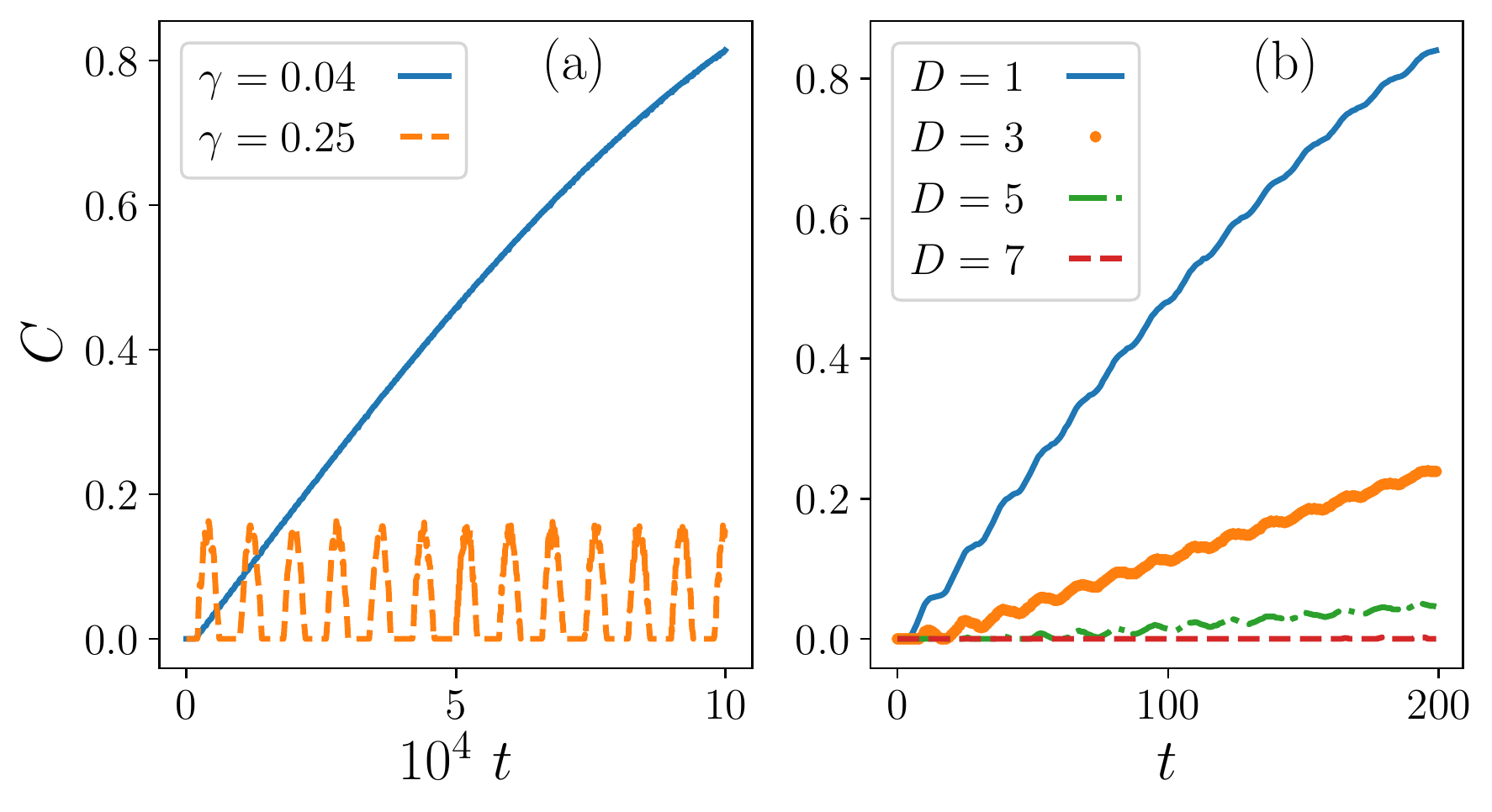}
\caption{Concurrence as a function of time (a) for different values of $\gamma$ when the qubits are at distance $D=10$ and (b) as a function of $D$ for $\gamma = 0.04$. 
The chain is composed by $N = 10^4$ oscillators with $\omega_0 = 0.45$ and $\beta = 10^5$. The qubits are initially in a separable, pure state with Bloch angles $\theta_a = \theta_b = \pi/4$ and $\phi_a = \phi_b = 0$. It can be observed that there is a finite time $t_{\text{gen}}$ after which entanglement is generated. }
\label{fig:moderate}
\end{figure}

We now analyse the features of entanglement as a function of the distance $D$. Figure \ref{fig:moderate}(a) displays the time evolution of the concurrence for $D=10$ and for two different values of $\gamma$. Because of the relatively large distance, the effective coupling strength of the function $\varphi$ is reduced, so that the dynamics determined by $\varphi$ is the one corresponding to the weak coupling regime: the concurrence oscillates at the frequency $\Omega_0$. Due to the different scaling with $D$ the effect of the attenuation becomes relevant at $\gamma=0.25$ and significantly reduces the maximal value that the concurrence can attain. At larger values of $\gamma$ we do not observe entanglement. 

Whenever we find entanglement, we observe that at short time the concurrence starts to grow after a finite time $t_{\rm gen}$ has elapsed, while for $t<t_{\rm gen}$ there is no entanglement between the qubits. 
Figure \ref{fig:moderate}(b) compares the concurrence at relatively short times and for different distances $D$, showing that $t_{\rm gen}$ increases monotonically with $D$. This result suggests that entaglement propagates at a finite velocity.

\begin{figure}
\includegraphics[scale= 0.5]{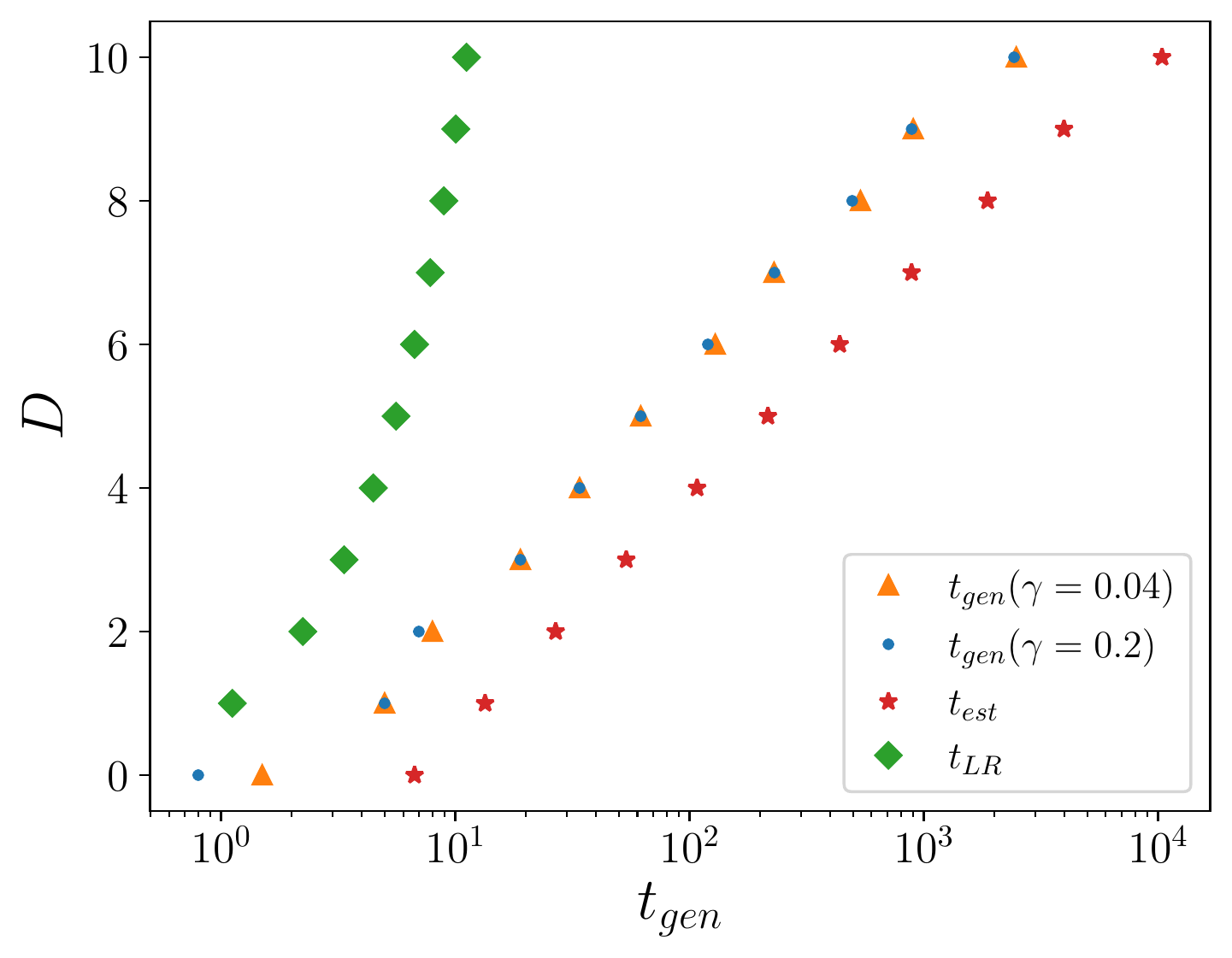}
\caption{Relation between the distance $D$ and the delay time $t_{\rm gen}$ after which the concurrence grows. $t_{\rm gen}$ is extracted from the numerical data and corresponds to the time for which the concurrence starts to grow continuously. The data have been evaluated for $\gamma = 0.04$ (orange triangle) and $\gamma = 0.1$ (blue dot). The other parameters are the same as in Fig. \ref{fig:moderate}. The red {stars} report the values predicted by the estimate of Eq.\ \eqref{Eq:est}. 
The green {diamonds} give the propagation time of the light cone with the Lieb-Robinson bound velocity $v_{LR}$ and are plotted for comparison. {The threshold was set such that $C(t_{\rm gen})=0.001$, corresponding to the maximal value that the concurrence reaches in the universal regime, where the decoherence function in the propagator scales with $t^3$. In this regime, in particular, $C$ has no monotonous behavior but fluctuates between 0 and 0.001.}} 
\label{fig:LR}
\end{figure}

Figure \ref{fig:LR} shows $t_{\rm gen}$ as a function of $D$. The time scale of entanglement generation is relatively independent of $\gamma$ and scales exponentially with the distance. We compare $t_{\rm gen}$ with the time scale $t_{LR}=D/v_{LR}$ that an excitation needs for covering the distance $D$ at the velocity of the Lieb-Robinson bound, Eq. \eqref{eq:LR}.  The comparison shows that $t_{\rm gen}$ is consistent with the Lieb-Robinson bound, but cannot be related to cone-light propagation across the chain. The dependence of $t_{\rm gen}$ on $D$ can be understood by considering that entanglement is generated by the collective Lamb shift. Therefore, it is generated at times where the coherent term in the imaginary part of the propagator, $\Omega_0(D)t$, exceeds the dephasing term.  For short times dephasing is dominant while for $t\gtrsim t_{\rm max}$ the dephasing term starts to oscillate with the upper bound $4\gamma^2/\omega_0^3$. We identify the time scale for entanglement generation using an equation that overestimates its value, $\Omega_0t_{\rm gen}= 4\gamma^2/\omega_0^3$. It gives
\begin{eqnarray}
\label{Eq:est}
	 t^{\rm est}_\mathrm{gen} \sim \frac{2\pi/\omega_0^3}{\int_0^{2\pi} {\rm d}k\frac{ \cos(kD)}{\omega(k)^{2}}} \,,
\end{eqnarray}  
where we have taken the continuum limit of Eq. \eqref{Eq:Omega:D}, $\Omega_0(D)\simeq \frac{4\gamma^2}{2\pi}\int_0^{2\pi} {\rm d}k\frac{ \cos(kD)}{\omega(k)^{2}}$. The resulting expression is independent of $\gamma$ and is reported in Fig. \ref{fig:LR}: $t^{\rm est}_\mathrm{gen}$ overestimates the time extracted from the numerical data but has the same functional dependence on the distance $D$. These considerations also clarify why we do not observe entanglement deep in the strong coupling regime for $D>0$. In this case, in fact, the density matrix decays to a statistical mixture at a faster time scale than $t_{\rm gen}$, or, in other words, the lifetime of entanglement is shorter than the time scale at which it can be generated. 

\begin{figure}
  \includegraphics[scale = 0.5]{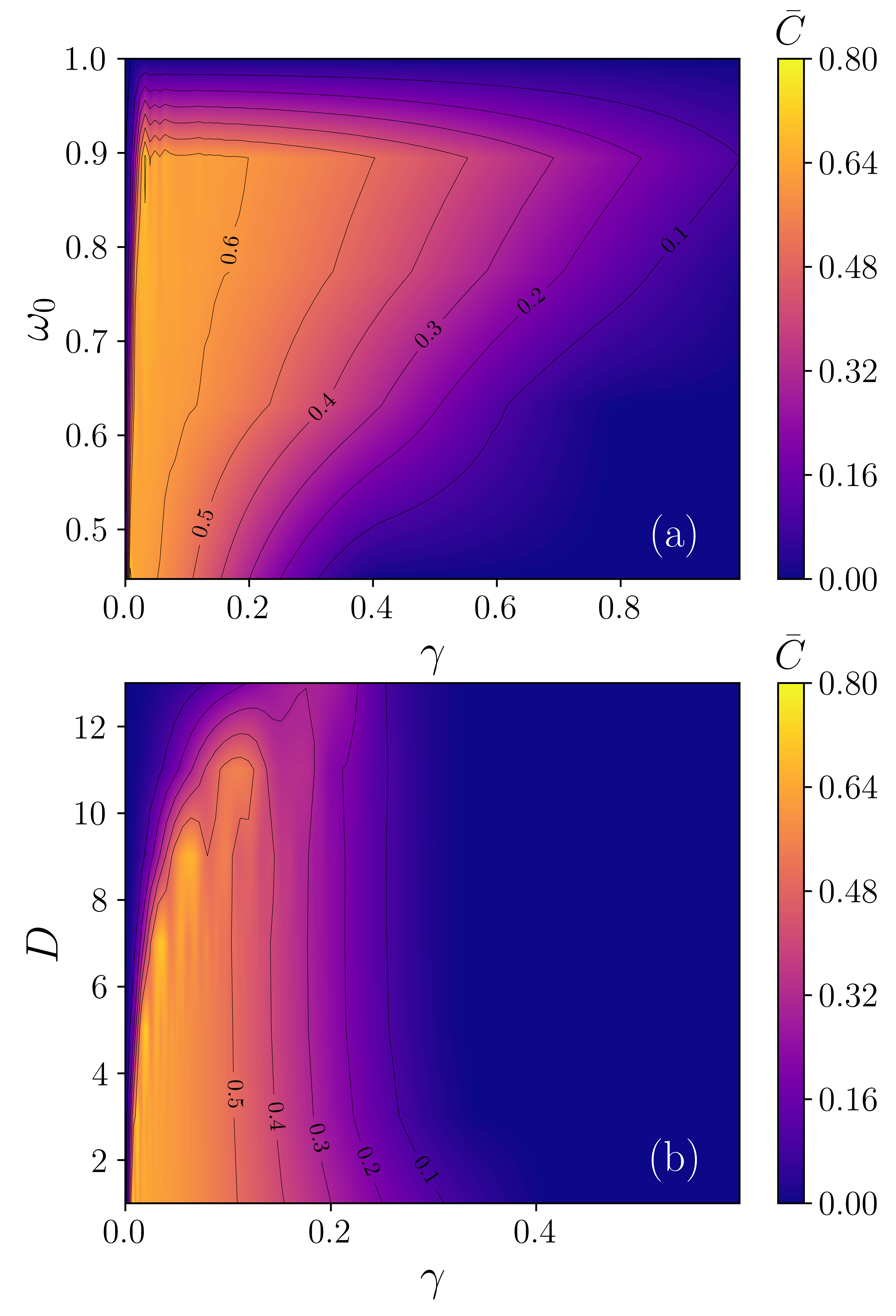}
  \caption{(color online) Color plot of the average Concurrence $\bar{C}$ (a) as a function of $\gamma$ and $\omega_0$ when the qubit are at distance $D=1$ and (b) as a function of the distance $D$ and of $\gamma$ for $J_0 = 0.8$ ($\omega_0=0.45$). The chain is composed by $N=10^4$ and is initially at inverse temperature $\beta = 10^5$. The two qubits are in a separable, pure state, Eq. \eqref{eq_bloch} with Bloch angles $\theta_a = \theta_b = \pi/4$ and $\phi_a = \phi_b = 0$. The concurrence was averaged over $t_{\rm end}=10^6$.}
\label{fig:phase_diag}
\end{figure}

Figure \ref{fig:phase_diag}(a) displays the average concurrence as a function of the coupling strength $\gamma$ and of the frequency $\omega_0$ when the qubits are at distance $D=1$. By comparing with 
Fig. \ref{fig:phase_diag_same}(a) for $D=0$, we observe the same qualitative behavior with two salient differences: (i) for the same value of $\gamma$ and $\omega_0$ the concurrence is generally smaller, which we attribute to the fact that $D>0$ leads to an effectively reduced coupling strength. Moreover, (ii)  the concurrence decreases to zero for $\omega_0\gtrsim 0.9$. This latter behavior is due to the fact that the band width becomes very small, $\Delta\omega\lesssim 0.1$, the band becomes flatter, and the collective Lamb shift, that  is essential for generating entanglement, vanishes as $\Omega_0\sim \int_0^{2\pi}{\mathrm{d} k\cos(2kD)}= 0 $.  

The average concurrence as a function of $D$ and $\gamma$ is shown in Fig.  \ref{fig:phase_diag}(b). Deep in the weak coupling regime we observe that entanglement is generated when $\gamma$ exceeds a threshold value that depends on $D$, and more specifically, it increases monotonically with the distance. This behavior  is a consequence of the scaling of $t_{\rm gen}$ with the distance $D$: In the region where the concurrence vanishes, $t_{\rm gen}$ is larger than the total evolution time (Simulations performed for smaller $t_{\rm end}$ show, in fact, that this threshold moves towards larger values of $\gamma$, see Appendix \ref{App:B}). Entanglement vanishes again at large $\gamma$, when the coupling strength reaches the strong coupling regime and damping suppresses quantum coherence. 

{
\subsection{Discussion}
The analysis so far is based on the properties of the propagator under the assumption that the symmetric and the antisymmetric subspaces are decoupled. We discuss now the case when this property, emerging from a symmetry of the Hamiltonian, is no longer fulfilled.  We now assume that the qubits couple with different coupling strengths, denoted by $\gamma_a$ and $\gamma_b$, to the chain oscillator(s). In this case the propagator for the symmetric and antisymmetric subspaces, separately, is the same as for the symmetric case, with now $\gamma=(\gamma_a+\gamma_b)/2$. Additionally, the two subspaces are now coupled with one another with strength $\Delta\gamma=(\gamma_a-\gamma_b)/2$. In order to assess the effect of the asymmetry, we compare different cases at constant $\gamma$ and varying $\Delta\gamma$.\\
In Appendix \ref{App:C} we provide some details of our study. The results show that, when the qubits couple to the same oscillator, then the analysis for $\Delta\gamma=0$ essentially applies also for $\Delta\gamma>0$ as long as $\Delta\gamma\ll\gamma$. In detail, in the weak coupling regime the entanglement is reduced by an amount proportional to $\Delta\gamma/\gamma$, while in the strong coupling regime the effects of the asymmetry $\Delta\gamma$ are irrelevant, since the corresponding time scale is longer than the lifetime of entanglement. When the qubits couple to different oscillators, the asymmetry can be neglected as long as the corresponding time scale is longer than the time scale $t_{\rm gen}$ at which entanglement is generated, and which depends solely on the chain spectral properties. }

\section{Conclusions}
\label{sec:5}

In this work we have analysed the dynamics of entanglement that is generated between two qubits coupled to a non-Markovian bath, here modelled by a chain of oscillators with a gapped spectrum. Using the exact solution for the qubits propagator we could unravel the processes that lead to entanglement between the two qubits and to its decay. Entanglement is generated {for a certain class of initial, separable states of the qbits} by coherent, Hamiltonian processes, which are reminiscent of the collective Lamb shift of dipolar systems. Dephasing mechanism lead to collapse and revival of the concurrence, while damping tends to suppress entanglement. Their interplay is controlled by the coupling strength of the qubits to the chain and by the elastic constant coupling the oscillators: Large coupling strengths tend to suppress entanglement, while instead a large elastic constant tends to favour it. {We emphasize that, in all situations here discussed, the environment is initially in a thermal state. Its capability to generate correlations requires that the initial temperature is below an upper bound, which our model allows to determine.}

Interestingly, entanglement is generated after a finite time has been elapsed. This behavior is due to the interplay between the effective coherent dynamics and the dephasing mechanism, the corresponding time scale is determined by the spectral properties of the chain's normal modes: entanglement is generated on time scales where the propagator does not exhibit the characteristic universal scaling with time. The scaling of the velocity, with which the qubits become entangled, is exponential with the qubit distance. This is consistent with the Lieb-Robinson bound, but cannot be related to light-cone propagation in the chain. This behaviour seems at odds with the picture developed by Calabrese and Cardy on the post-quench dynamics of entanglement as a consequence of ballistically propagating quasi-particles \cite{Calabrese_2009}. Indeed, entanglement is generated by the effect of the collective Lamb shift, introducing an additional characteristic frequency scale in the problem. Moreover, the geometry of the coupling between chain and qubit introduces a characteristic wavelength scaling with the qubits distance $D$ that has a similar effect as the free spectral range of a Fabry-Perot resonator \cite{Berman:1994,Wolf:2011}.  

Our findings provide important guidelines for designing quantum steering protocols \cite{Roy:2020} which go beyond the Markovian paradigm of projective measurements \cite{Alcalde:2023}.
The configuration here considered could be implemented in quantum optics experiment: the environment would be mimicked by coupled microcavities \cite{Tanese:2013,Wachter:2019}, by an optomechanical array \cite{Eichenfield:2009,Ludwig:2013}, or by the transverse modes of an ion chain  \cite{Porras:2004,Serafini:2009}. For the dynamics here discussed the qubits frequencies shall be the smallest frequency scale of the problem. Entanglement between the qubits could be detected using the protocol discussed in Ref. \cite{Serafini:2009,Taketani:2014} or the measures discussed in Ref. \cite{Cattaneo:2021}. 

The dynamics of entanglement shall change substantially for a gapless spectrum, where the weight of damping and dephasing terms in general increases with time. Future works shall address how the velocity of entanglement propagation behaves in the presence of disorder \cite{Burrell:2007}, and how it is modified when the oscillators interact with long-range interactions, where the bound on the propagation of information follows a different scaling \cite{Tran:2020}.

{To conclude, we have analysed the entanglement generated between two qubits by a non-Markovian quantum channel and characterised the mechanisms that determine its onset, stability and/or decay. Our analysis can be extended to determine quantum information scrambling in the chain, by considering a qubit chain where information is mediated by the non-Markovian environment and the entanglement entropy in order to analyse the dynamics of correlations \cite{Shor:PRX}.}
. 

\section*{Acknowledgments}
The authors are grateful to E. Kajari, E. King, D. Karevski, F. Marquardt, B. Taketani, and P. Wendenbaum for discussions. 
This work was funded by the Deutsche Forschungsgemeinschaft (DFG, German Research Foundation), Project-ID 429529648 TRR 306 QuCoLiMa("Quantum Cooperativity of Light and Matter") and by the Bundesministerium f\"ur Bildung und Forschung (BMBF, Project "NiQ: Noise in Quantum Algorithms").


\appendix

\section{Correlation functions}
\label{App:A}

Below we report  the expectation value of the Pauli operators and of the correlation functions at equal time and at different times. These can be straightforwardly evaluated using the propagator. We compute the correlations with the reduced density matrix in the tensor product basis $\{\ket{++}_x,\ket{+-}_x,\ket{-+}_x,\ket{--}_x \}$ (labelled from $1$
to $4$ respectively for notation purpose). 

The time evolution for the expectation values of the Pauli matrices are:
\begin{eqnarray}
	\langle \sigma_x^a \rangle  &=& \rho_{11}(0) + \rho_{22}(0) - \rho_{33}(0) - \rho_{44}(0) \nonumber\\
	\langle \sigma_y^a \rangle  &=& 2 \,\text{Re}[\rho_{13}(0) e^{i(\phi^S(t) - \phi^A(t))}]\, e^{-f^S(t) - f^A(t)} \nonumber \\ &+&  2\,\text{Re}[\rho_{24}(0) e^{-i(\phi^S(t) - \phi^A(t))}]\, e^{-f^S(t) - f^A(t)}  \nonumber \\
	\langle \sigma_z^a \rangle  &=& 2 \, \text{Im}[\rho_{13}(0) e^{i(\phi^S(t) - \phi^A(t))}]\, e^{-f^S(t) - f^A(t)} \nonumber \\ &+& 2 \, \text{Im}[\rho_{24}(0) e^{-i(\phi^S(t) - \phi^A(t))}] \, e^{-f^S(t) - f^A(t)} 
\end{eqnarray}

The correlation functions at equal time read:
\begin{eqnarray}
	\langle \sigma_x^a \sigma_x^b \rangle &=& \rho_{11}(0) - \rho_{22}(0) -\rho_{33}(0) + \rho_{44}(0)  \nonumber \\
	\langle \sigma_y^a \sigma_y^b \rangle &=& 2\, \text{Re}[\rho_{14}(0)]\, e^{-4 f^S(t)} + 2\, \text{Re}[\rho_{23}(0)]\,e^{-4 f^A(t)}  \nonumber\\
	\langle \sigma_z^a \sigma_z^b \rangle &=&  -2\, \text{Re}[\rho_{14}(0)]\, e^{-4 f^S(t)} + 2\, \text{Re}[\rho_{23}(0)]\,e^{-4 f^A(t)}   \nonumber\\
	\langle \sigma_x^a \sigma_y^b \rangle  &=& 2 \,\text{Re}[\rho_{12}(0) e^{i(\phi^S(t) - \phi^A(t))}] \, e^{-f^S(t) - f^A(t)} \nonumber \\  &-& 2 \,\text{Re}[\rho_{34}(0) e^{-i(\phi^S(t) - \phi^A(t))}] \, e^{-f^S(t) - f^A(t)} \nonumber\\
	\langle \sigma_x^a \sigma_z^b \rangle  &=& 2 \, \text{Im}[\rho_{12}(0) e^{i(\phi^S(t) - \phi^A(t))}]\, e^{-f^S(t) - f^A(t)}  \nonumber \\ &-& 2 \, \text{Im}[\rho_{34}(0) e^{-i(\phi^S(t) - \phi^A(t))}] e^{-f^S(t) - f^A(t)} \nonumber\\
	\langle \sigma_y^a \sigma_z^b \rangle  &=& 2\, \text{Im}[\rho_{14}(0)]\, e^{-4 f^S(t)} - 2\, \text{Im}[\rho_{23}(0)]\,e^{-4 f^A(t)} \nonumber\\
\end{eqnarray}

\begin{figure}[!htpb]
	\includegraphics[scale =0.5]{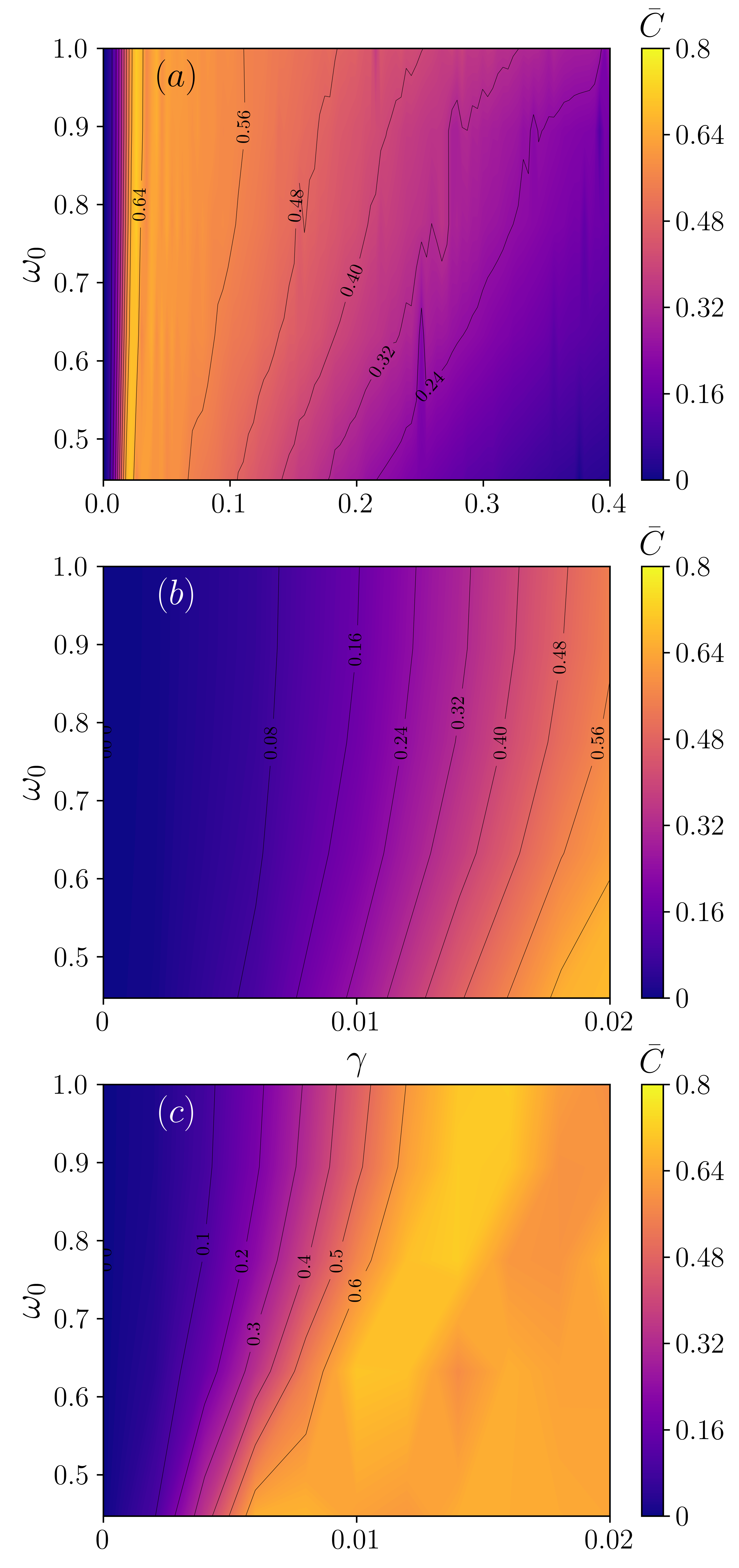}
	\caption{(color online) Same as Fig. \ref{fig:phase_diag_same}(a) but for $t_{\rm end}= 10^3$. Subplot (b) shows a zoom into the region at small $\gamma$. Subplot (c) shows the zoom in the same region for the propagation time of Fig. \ref{fig:phase_diag_same}(a).}
	\label{fig:phase_diag_same:3}
\end{figure}

{The correlation function at different times can be computed by first evolving the density matrix $(\,\rho_t = U(t) \rho_0 U^{\dagger}(t) \,)$ for time $t$, and then modifying it accordingly. We denote the modified density matrix as $\tilde{\rho}^{\alpha}(t) = \rho_t (\,\sigma_\alpha^a \otimes \mathds{1}\,) $ and let $t^{\prime} =  t + \tau$, then the correlation functions have the form:}

	\begin{eqnarray}
	\langle \sigma_x^a(t) \sigma_x^b(t^{\prime}) \rangle  &=& \tilde{\rho}^x_{11}(t)  - \tilde{\rho}^x_{22}(t) + \tilde{\rho}^x_{33}(t) - \tilde{\rho}^x_{44}(t) \nonumber\\
	\langle \sigma_y^a(t) \sigma_y^b(t^{\prime}) \rangle &=& 2 \,\text{Re}[ \tilde{\rho}^y_{12}(t) e^{i(\phi^S(\tau) - \phi^A(\tau))}] e^{-f^S(\tau) - f^A(\tau)}  \nonumber \\
	&+&  2\text{Re}[\tilde{\rho}^y_{34}(t) e^{-i(\phi^S(\tau) - \phi^A(\tau))}] e^{-f^S(\tau) - f^A(\tau)} \nonumber\\
	\langle \sigma_x^a(t) \sigma_y^b(t^{\prime}) \rangle &=&  2 \,\text{Re}[ \tilde{\rho}^x_{12}(t) e^{i(\phi^S(\tau) - \phi^A(\tau))}] e^{-f^S(\tau) - f^A(\tau)} \nonumber \\
	&+&  2\text{Re}[ \tilde{\rho}^x_{34}(t) e^{-i(\phi^S(\tau) - \phi^A(\tau))}] e^{-f^S(\tau) - f^A(\tau)} \nonumber\\
	\langle \sigma_z^a (t) \sigma_z^b(t^{\prime}) \rangle &=& 2 \,\text{Im}[\tilde{\rho}^z_{12}(t) e^{i(\phi^S(\tau) - \phi^A(\tau))}] e^{-f^S(\tau) - f^A(\tau)} \nonumber \\
	&+&  2 \text{Im}[\tilde{\rho}^z_{34}(t) e^{-i(\phi^S(\tau) - \phi^A(\tau))}] e^{-f^S(\tau) - f^A(\tau)} \nonumber\\
	\langle \sigma_x^a(t) \sigma_z^b(t^{\prime}) \rangle &=& 2 \, \text{Im}[\tilde{\rho}^x_{12}(t) e^{i(\phi^S(\tau) - \phi^A(\tau))}] e^{-f^S(\tau) - f^A(\tau)}  \nonumber \\
	&+& 2 \text{Im}[\tilde{\rho}^x_{34}(t) e^{-i(\phi^S(\tau) - \phi^A(\tau))}] e^{-f^S(\tau) - f^A(\tau)} \nonumber\\
	\langle \sigma_y^a(t) \sigma_z^b(t^{\prime}) \rangle &=& 2 \,\text{Im}[\tilde{\rho}^y_{12}(t) e^{i(\phi^S(\tau) - \phi^A(\tau))}] e^{-f^S(\tau) - f^A(\tau)}\nonumber \\
	&+&  2 \text{Im}[\tilde{\rho}^y_{34}(t) e^{-i(\phi^S(\tau) - \phi^A(\tau))}]  e^{-f^S(\tau) - f^A(\tau)} \nonumber \\
\end{eqnarray}


\section{Average concurrence for $t_{\rm end}=10^3$}
\label{App:B}

We report the average concurrence for $t_{\rm end}=10^3$, which for the considered chain size is below the Poincar\'e time. Figure \ref{fig:phase_diag_same:3}(a) shall be compared with Fig. \ref{fig:phase_diag_same}(a): apart for larger fluctuations, the qualitative behavior is the same. Subplot (b) and (c) show the zoom in the region at small $\gamma$ for $t_{\rm end}=10^3$ and $t_{\rm end}=10^6$, respectively: For longer integration times entanglement is generated at smaller values of $\gamma$.

\begin{figure}[!htpb]
	\includegraphics[scale = 0.5]{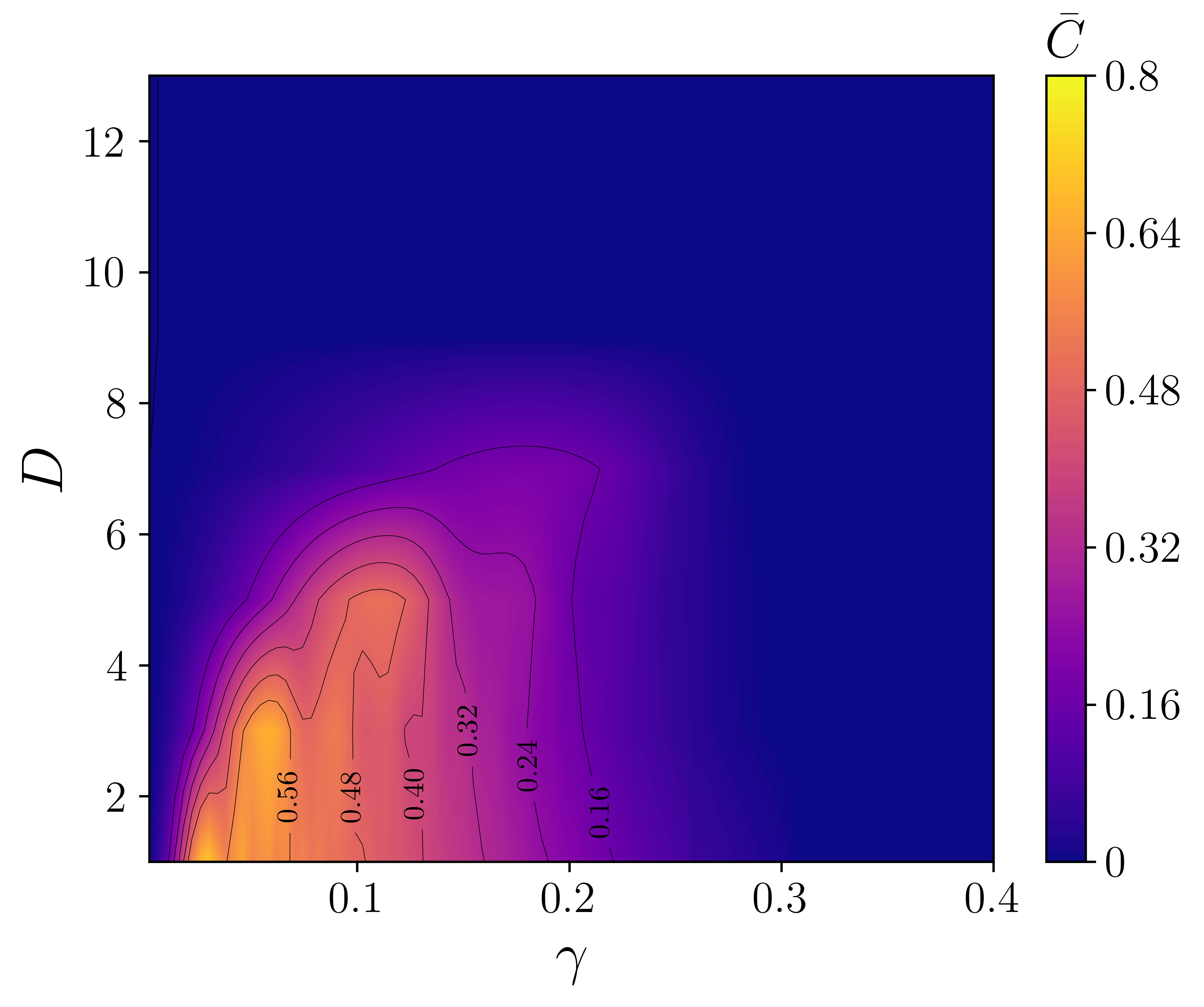}
	\caption{(color online) Same as Fig. \ref{fig:phase_diag}(b) but for $t_{\rm end}= 10^3$.}
	\label{fig:phase_diag:3}
\end{figure}

Figure \ref{fig:phase_diag:3} is calculated for the same parameter of Fig. \ref{fig:phase_diag}(b) but integration time  $t_{\rm end}=10^3$: entanglement is established till distance $D\sim 6$ due to the exponential scaling with the distance of the velocity with which it propagates along the chain. 

\section{Effect of Asymmetric Coupling Strength in the Interaction Hamiltonian}
\label{App:C}

In the main text we have used the uniform coupling strength for the two qubits in the interaction Hamiltonian \eqref{eq:H-int}, however one may also ask how the results modify for the case of asymmetric coupling strength. In order to analyze the role of asymmetry, we introduce the new interaction Hamiltonian

\begin{equation}
 H_{\text{int}} =  - \hbar \left( \bar{\gamma}_a \sigma_x^a x_{l_a} + \bar{\gamma}_b \sigma_x^b x_{l_b} \right).
\end{equation}

Following the same treatment as in the main text, the interaction Hamiltonian can be written in the normal coordinates as

\begin{eqnarray}
    H_{\text{int}} = - \sum_n \Big( && \frac{\tilde{\gamma}^S_{a,n} + \tilde{\gamma}^S_{b,n}}{2} \, \tilde{x}_n^S (\sigma_x^a + \sigma_b^x)/2 \nonumber \\
    &+& \frac{ \tilde{\gamma}^A_{a,n} + \tilde{\gamma}^A_{b,n} }{2} \, \tilde{x}_n^A (\sigma_x^a - \sigma_b^x)/2 \nonumber \\
    &+&\frac{ \tilde{\gamma}^A_{a,n} - \tilde{\gamma}^A_{b,n} }{2} \, \tilde{x}_n^A (\sigma_x^a + \sigma_b^x)/2 \nonumber \\ 
    &+& \frac{ \tilde{\gamma}^S_{a,n} - \tilde{\gamma}^S_{b,n} }{2} \, \tilde{x}_n^S (\sigma_x^a - \sigma_b^x)/2 \Big),
\end{eqnarray}

where the coupling constant reads as \eqref{eq:gamma:D} with $\gamma$ replaced with $\gamma_j$, with $j =a, b$. One can easily recover the case of same coupling strength \eqref{eq:H_int-normal} by using the condition $\gamma_a = \gamma_b = \gamma$. In the interaction Hamiltonian, the first two terms is analogous to the case of uniform coupling strength. For understanding the role of asymmetry, the last two terms containing the difference between coupling strength have to be analyzed. For the same we keep the sums $ \gamma_{+}^{S,A} \equiv \tilde{\gamma}^{S,A}_{a,n} + \tilde{\gamma}^{S,A}_{b,n}$ constant and vary the differences $ \gamma_{-}^{S,A} \equiv \tilde{\gamma}^{S,A}_{a,n} - \tilde{\gamma}^{S,A}_{b,n}$. Following the same recipe as discussed in  \cite{Braun:2001,Wendenbaum:2020}, the reduced density matrix elements in the same subspace evolves as

\begin{eqnarray}
		&&\bra{b_i^{\alpha}}\rho_{d}(t)\ket{b_j^{\alpha}}  = \bra{b_i^{\alpha}}\rho_{d}(0)\ket{b_j^{\alpha}}    \label{eq:evolution_asym} \\
		&& \times  \exp\left(- [f_+^{\alpha}(t) + f_-^{\beta}(t) ]\left(b_i^{\alpha}-b_j^{\alpha}\right)^2\right)\,,\nonumber
\end{eqnarray}
where $\alpha, \beta = S,A$, $\alpha \neq \beta$ and

\begin{align}
f_{\pm}^\alpha(t) &=\frac{1}{2} \sum_n \frac{\tilde{\gamma}_{n,\pm}^{\alpha\,2}}{\omega_n^{\alpha\,3}}\left(2\bar n(\omega_n^\alpha) + 1\right)\left(1-\cos\left(\omega_n^\alpha t \right)\right) \,, \label{eq:fsa_asym}
\end{align}
with the mean thermal occupation of the $\alpha$-th symmetric (antisymmetric) mode of the chain being given by the Bose-Einstein statistics $\bar n(\omega)  = 1 / (e^{\beta\omega} - 1 )$. The dynamics leads to additional damping (with respect to uniform coupling strength) of the off-diagonal elements within the same subspace which scales linearly with the differences $\gamma_{-}^{S,A}$ , while the diagonal elements are still constant in time.

\begin{figure}
	\includegraphics[scale = 0.5]{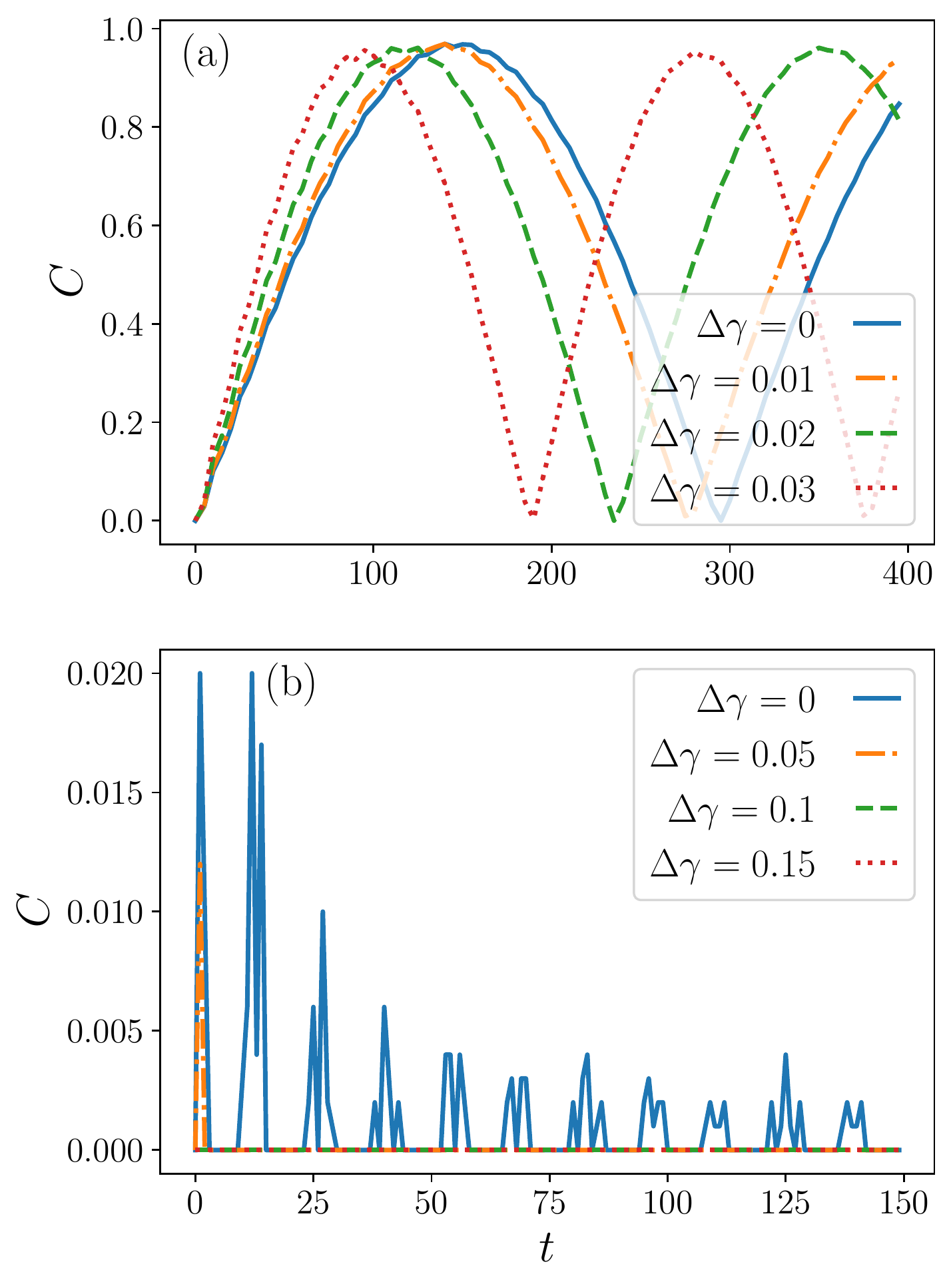}
	\caption{Concurrence $C$ for different values of difference in coupling strength $\Delta \gamma$ while keeping the sum $\gamma_a + \gamma_b)/2 $ fixed for the case when qubits are coupled to the same oscillator. Subplot (a) shows the dynamics of concurrence in the weak coupling regime, $(\gamma_a + \gamma_b)/2 \ll \omega_0^{3/2}$  . Here we chose $(\gamma_a + \gamma_b)/2 = 0.04$. The plot for the concurrence dynamics in strong coupling regime ($(\gamma_a + \gamma_b)/2 \gg \omega_0^{3/2}$) is shown in subplot (b) with $(\gamma_a + \gamma_b)/2 = 0.6$. The qubits are initially prepared in the same pure state \eqref{eq_bloch} with $\theta_a = \theta_b = \pi/4$ and $\phi_a = \phi_b = 0$. The chain is composed by $N=10^4$ oscillators with $\omega_0=0.45$ and inverse temperature $\beta = 10^5$.}
	\label{fig:asym:1}
\end{figure}

The time evolution of the matrix elements between the symmetric and the antisymmetric subspaces takes instead the form:

\begin{eqnarray}
\bra{b_i^{\alpha}}\rho_{d}(t)\ket{b_j^{\beta}}_{\beta\neq\alpha}  &=& \bra{b_i^{\alpha}}\rho_{d}(0)\ket{b_j^{\beta}}_{\beta\neq\alpha}   \label{eq:evolution:off_asym} \\
\times \exp \Bigg( &-& f_+^{\alpha}(t)-f_-^{\alpha}(t) - f_+^{\beta}(t) - f_-^{\beta}(t)  \nonumber \\ &-&  (-1)^{i+j} [F^\alpha + F^\beta]\nonumber \\ &+& i(\varphi_+^{\alpha}(t) + \varphi_-^{\alpha}(t)  - \varphi_+^{\beta}(t)) - \varphi_-^{\beta}(t)\Bigg)\,, \nonumber
\end{eqnarray}
where the time-dependent phases read  
\begin{align}
\varphi_{\pm}^\alpha(t) &= \frac{1}{2}\sum_n\left(\frac{\tilde{\gamma}_{n,\pm}^{\alpha\,2}}{\omega_n^{\alpha\,2}}t-\frac{\tilde{\gamma}_{n,\pm}^{\alpha\,2}}{\omega_n^{\alpha\,3}}\sin\left(\omega_n^{\alpha}t\right)\, \right)\,.\label{eq:phisa_asym}
\end{align}
and the additional decay term

\begin{align}
F^\alpha(t) &=\frac{1}{2} \sum_n \frac{\tilde{\gamma}_{n,+}^{\alpha} \tilde{\gamma}_{n,-}^{\alpha}}{\omega_n^{\alpha\,3}}\left(2\bar n(\omega_n^\alpha) + 1\right)\left(1-\cos\left(\omega_n^\alpha t \right)\right) \,, \label{eq:F_asym}
\end{align}

\begin{figure}
	\includegraphics[scale = 0.5]{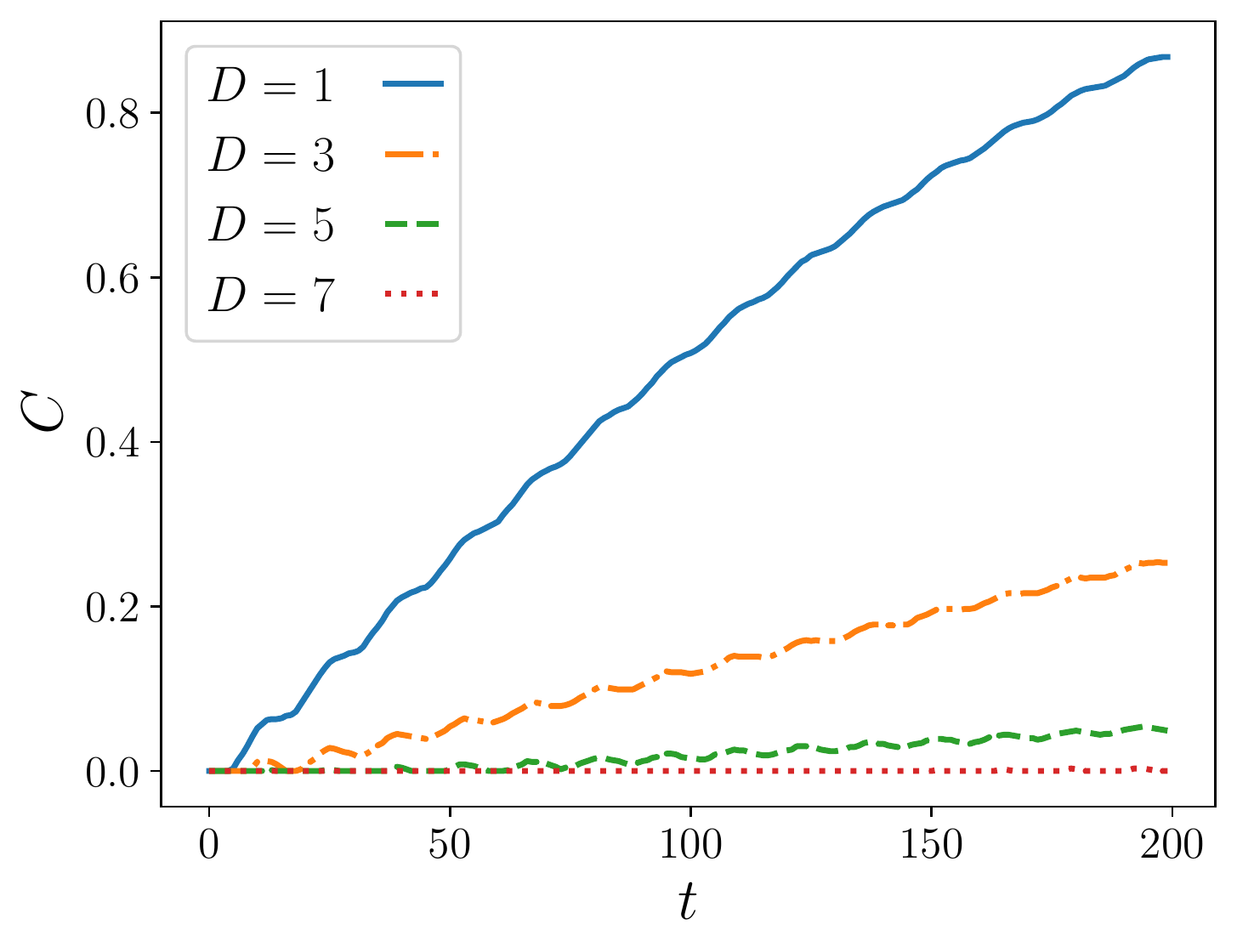}
	\caption{Concurrence as a function of time for different distances with constant different coupling strength. Here, $\gamma_a = 0.05$, $\gamma_b =0.03$. The qubits are initially prepared in the same pure state \eqref{eq_bloch} with $\theta_a = \theta_b = \pi/4$ and $\phi_a = \phi_b = 0$. The chain is composed by $N=10^4$ oscillators with $\omega_0=0.45$ and inverse temperature $\beta = 10^5$.}
	\label{fig:asym:2}
\end{figure}

The same site dynamics can be recovered by putting all the antisymmetric coupling vector to 0. In contrast to the uniform strength case, here the antisymmetric subspace is not protected from decoherence due to difference in coupling strength. The damping in the antisymmetric subspace decays linearly with the difference. 

One can identify the same coupling regimes as before with a modified coupling strength. The weak coupling regime is when $(\gamma_a + \gamma_b)/2 \ll \omega_0^{3/2}$. This regime is mainly governed by coherent oscillations with negligible decoherence and damping. The strong coupling regime is the opposite limit when both decoherence and damping effects become dominant that is $(\gamma_a + \gamma_b)/2 \gg \omega_0^{3/2}$. In this framework, we plot below the dynamics of entanglement by keeping the sum $(\gamma_a + \gamma_b)/2$ fixed and vary the difference $\Delta \gamma = |\gamma_a - \gamma_b|/2$. 

For the case of $D = 0$, we observe in Fig. \ref{fig:asym:1}(a), that in the weak coupling regime the difference in coupling strength benefits the concurrence to grow quicker and this is due to the additional frequency in the collective Lamb shift $\Omega_0$  which scales with $\Delta \gamma$. For the case of strong coupling regime Fig. \ref{fig:asym:1}(b) , we observe additional decay of concurrence maxima, but it retains the collapse and revival type of dynamics. 

As a function of distance, the entanglement generation time have the same behavior even with the difference in coupling strength  which is clearly visible in Fig. \ref{fig:asym:2}.


\bibliography{reference}

\end{document}